\documentclass[preprint]{aastex}
\usepackage{graphicx,natbib,amsmath, apjfonts,color, longtable}

\newcommand{\dtwo}{{$_2$}}
\newcommand{\dthree}{{$_3$}}
\newcommand{\jon}{{$^+$}}

\newcommand{\pHtre}{$p_{3}$}
\newcommand{\pHtva}{$p_{2}$}

\slugcomment{To appear in ApJ}
\onecolumn

\shorttitle{First time-dependent study of H$_2$ and H$_3^+$ \textit{ortho}$-$\textit{para} chemistry in the diffuse ISM}
\shortauthors{Albertsson et al. }
\begin{document}

\title{First time-dependent study of H$_2$ and H$_3^+$ \textit{ortho}$-$\textit{para} chemistry in the diffuse interstellar medium{: observations meet theoretical predictions}\footnote{Partly based on observations collected at the European Organization for Astronomical Research in the Southern Hemisphere, Chile, as part of program 088.C-0351}}

\author{T. {Albertsson}\altaffilmark{1}, N. {Indriolo}\altaffilmark{2}, H. {Kreckel}\altaffilmark{3}, D. {Semenov}\altaffilmark{1}, K.~N. {Crabtree}\altaffilmark{4} and Th. {Henning}\altaffilmark{1}}
\affil{(1) Max-Planck-Institut f\"ur Astronomie, K\"onigstuhl 17, 69117 Heidelberg, Germany \\
(2) Department of Physics and Astronomy, Johns Hopkins University, Baltimore, MD 21218, USA \\
(3) Max-Planck-Institut f\"ur Kernphysik, 69117 Heidelberg, Germany \\
(4) Harvard--Smithsonian Center for Astrophysics, 60 Garden St., Cambridge, MA 02138, USA}

\begin{abstract}
The chemistry in the diffuse interstellar medium initiates the gradual increase of molecular complexity during the life cycle of matter. A key molecule that enables build-up of new molecular bonds and new molecules via proton-donation is H$_3^+$. Its evolution is tightly related to molecular hydrogen and thought to be well understood. However, recent observations of \textit{ortho} and \textit{para} lines of H$_2$ and H$_3^+$ in the diffuse ISM showed a puzzling discrepancy in nuclear spin excitation temperatures and populations between these two key species. H$_3^+$, unlike H$_2$, seems to be out of thermal equilibrium, contrary to the predictions of modern astrochemical models. We conduct the first time-dependent modeling of the \textit{para}-fractions of H$_2$ and H$_3^+$ in the diffuse ISM and compare our results to a set of line-of-sight observations, including new measurements presented in this study. We isolate a set of key reactions for H$_3^+$ and find that the destruction of the lowest rotational states of H$_3^+$ by dissociative recombination largely control its \textit{ortho}/\textit{para} ratio. A plausible agreement with observations cannot be achieved unless {a ratio larger than 1:5 for} the destruction of $(1,1)-$ and $(1,0)-$states of H$_3^+$ is assumed. Additionally, {an increased CR ionization rate to $10^{-15}$ s$^{-1}$ further improves the fit whereas} variations of other individual physical parameters, such as density and chemical age, have only a minor effect on the predicted \textit{ortho}/\textit{para} ratios. Thus our study calls for new laboratory measurements of the dissociative recombination rate and branching ratio of the key ion H$_{3}^{+}$ under interstellar conditions.
\end{abstract}


\section{Introduction}
H\dthree\jon~plays a pivotal role in the gas-phase chemistry of the interstellar medium due to its very low proton affinity, allowing it to transfer a proton to many neutral atoms and molecules (exception being N and O\dtwo). The chemistry of H\dthree\jon~is straightforward. The formation process via the ion-molecule reaction H$_{2}^{+}$ + H$_{2}$ -> H$_{3}^{+}$ + H is well-established \citep[e.g.][]{2006RSPTA.364.3049L}. The destruction of H$_{3}^{+}$ can occur via ion-molecule reactions or dissociative recombination (DR) with free electrons. {Under typical conditions of different ISM environments H$_{3}^{+}$ was long believed to exist below observable limits. Still,} H\dthree\jon~was observed in the interstellar medium by \citet{1996Natur.384..334G} for the first time, followed by other detections \citep[e.g. ][]{1998Sci...279.1910M, 1999ApJ...510..251G, 2002ApJ...567..391M,2008ApJ...688..306G, 2007ApJ...671.1736I, 2012ApJ...745...91I}. These observations have also revealed several unexpected results, which are summarized below. 

In the simple gas-phase chemistry of H\dthree\jon~only three parameters can strongly affect its {steady-state} abundance: the dissociative recombination rate coefficients, the electron abundance, and the cosmic-ray (CR) ionization rate \citep{2003Natur.422..500M}. The former two parameters are thought to be well constrained in diffuse interstellar clouds \citep{1996ApJ...467..334C, 2003Natur.422..500M}, which leaves the CR ionization rate as the only controlling parameter. \citet{2003Natur.422..500M}, \citet{2007ApJ...671.1736I} and \citet{2012ApJ...745...91I} observed absorption lines of H\dthree\jon~toward several diffuse cloud sight lines, and inferred CR ionization ratios  on the order of $\sim 10^{-16}$ s$^{-1}$, about an order of magnitude higher than the value inferred for dark prestellar cores \citep[$\sim 10^{-17}$ s$^{-1}$, see e.g.][]{1998ApJ...506..329W, 2000A&A...358L..79V, 2003Ap&SS.285..619C, 2006RSPTA.364.3101V}. {If one relaxes the steady-state approximation, the density starts to play an important role in the H$_3^+$ evolution \citep[see e.g.][]{2009ApJ...706.1429C, 2013IAUS..292..223F}. }

{Furthermore}, observations of the average excitation temperature derived from the two lowest rotational states of H\dthree\jon, $T($H$_3^+$) $\approx 30$ K \citep{2007ApJ...671.1736I, 2012ApJ...745...91I}. It differs significantly from that of the two lowest rotational states of H\dtwo, $T_{01} \approx 70$ K \citep{2002ApJ...577..221R, 2009ApJS..180..125R}. Because the conversion between the two lowest nuclear spin states of H$_{2}$ in collisions with free protons is very efficient, the H$_{2}$ \textit{ortho}$-$\textit{para} ratio is expected to be thermalized with the gas kinetic temperature. Hence, the excitation temperature derived from the relative intensities of H\dtwo~\textit{ortho} and \textit{para} levels are also expected to be an accurate measure of the gas kinetic temperature in the diffuse ISM ($\approx 70$ K). Assuming that collisional thermalization between H$_{3}^{+}$ and H$_{2}$ is also efficient, in previous studies by \citet{2003Natur.422..500M} and \citet{2010ApJ...715..757G} the nuclear spin states of H$_{3}^{+}$ were assumed to be in thermal equilibrium with the kinetic cloud temperature. However, in later studies the excitation temperatures of H\dtwo~and H\dthree\jon directly derived from observations did not agree with each other, indicating that a large population of H$_3^+$ is not thermalized with the diffuse ISM gas.

\citet{2011ApJ...729...15C} have investigated this discrepancy by comparing observations of the nuclear spin temperature of H\dthree\jon~\citep[subsequently refined by][]{2012ApJ...745...91I} to that of H\dtwo~for a sample of diffuse interstellar clouds. Their results confirmed that the excitation temperature of H\dthree\jon~and H\dtwo~do not agree. \citet{2011ApJ...729...15C} concluded that the H\dthree\jon~\textit{ortho}/\textit{para} ratio is likely governed by a competition between the collisionally-driven thermalization of H\dthree\jon~and the DR reactions with electrons. The thermalization reaction \textit{ortho}/\textit{para}$-$H$_{3}^{+}$ + H$_{2}$ $\rightarrow$ \textit{ortho}$/$\textit{para} H$_{3}^{+}$ + H$_{2}$ has recently been studied experimentally by \citet{2012ApJ...759...21G}. It was found that the reaction has indeed the expected thermal outcome.

From a theoretical point of view, there are more unknown factors related to the chemistry of H\dthree\jon. Theoretical calculations have shown that the photodissociation of H\dthree\jon~is not efficient in the diffuse interstellar medium \citep[see e.g.][]{1987IAUS..120...51V}. In the absence of other abundant molecules like CO and H$_2$O, this leaves DR as the only major destruction pathway for H\dthree\jon. Recent theoretical calculations by \citet{santos:124309} predict the DR rate coefficient for \textit{para}$-$H$_{3}^{+}$ at low temperature to be an order of magnitude higher than that for \textit{ortho}$-$H$_{3}^{+}$. This claim has later been backed up by the plasma experiments of \citet{2011PhRvL.106t3201V}. Meanwhile, other laboratory groups have observed a different dependence, with similar dissociation rates between the two nuclear spin states of H$_{3}^{+}$ \citep[see e.g.][]{kreckel05,tom09, kreckel10}. 

In this paper we conduct the first time-dependent study of the \textit{ortho}$-$\textit{para} chemistry of H$_{3}^{+}$ in the diffuse interstellar medium. We isolate a set of key processes for the evolution of the \textit{ortho}- and \textit{para}-states of H$_3^+$, and we present new observational measurements to better test our model predictions. The paper is structured as follows. In Section~\ref{sec:obs} we discuss the new observations. In Section~\ref{sec:model} we describe the chemical and physical models utilized in the analysis of the observations. Our results and the underlying chemistry is presented and discussed in Section~\ref{sec:results}, followed by conclusions given in Section~\ref{sec:conclusions}. 

\section{Observations}\label{sec:obs}
{Following the initial comparison of the H$_3^+$ and H$_2$ \textit{ortho}/\textit{para} ratios presented by \citet{2011ApJ...729...15C} we proposed new observations of H$_3^+$ in eight diffuse cloud sight lines with measured H$_2$ column densities.  The intent of these new observations was to significantly expand the sample size from the five sight lines considered by \citet{2011ApJ...729...15C}. Targeted sight lines were selected based on large observed H$_2$ column densities \citep[$N({\rm H}_2)>10^{20}$~cm$^{-2}$][]{2002ApJ...577..221R, 2009ApJS..180..125R, 1977ApJ...216..291S}, and bright infrared background sources ($L>7.5$~mag) to maximize the likelihood of H$_3^+$ detections with relatively short ($\lesssim2$~hr) exposure times.  Despite these considerations, H$_3^+$ absorption lines were detected in only three of the eight targeted sight lines.  As the new H$_3^+$ detections are relevant to the current study, they are presented herein, whereas the non-detections will be presented in a future publication.}

The sight lines toward HD~27778, HD~43384, and HD~41117 were observed on 2011~Nov 6, 2011~Dec~1, and 2012~Apr~1, respectively, using the Cryogenic High-resolution Infrared Echelle Spectrograph \citep[CRIRES; ][]{2004SPIE.5492.1218K} on UT1 at the Very Large Telescope. Observations were performed in service mode, and CRIRES was used with its 0\farcs2 slit to provide a resolving power (resolution) of about 100,000 (3~km~s$^{-1}$), and a reference wavelength of 3715.0~nm to position the H$_3^+$~$R(1,1)^l$ ($\lambda=3.715479$~$\mu$m) transition on detector 3, and the $R(1,1)^u$ ($\lambda=3.668083$~$\mu$m) and $R(1,0)$ ($\lambda=3.668516$~$\mu$m) transitions on detector 1. The adaptive optics system was utilized in all cases to maximize starlight passing through the narrow slit. Spectra were obtained in an ABBA pattern with 10\arcsec\ between the two nod positions and $\pm$3\arcsec\ jitter width. Total integration times for the three targets were as follows: HD~27778: 34~min; HD~43384: 12 min; HD~41117: 12 min.

\begin{figure}
\epsscale{0.8}
\plotone{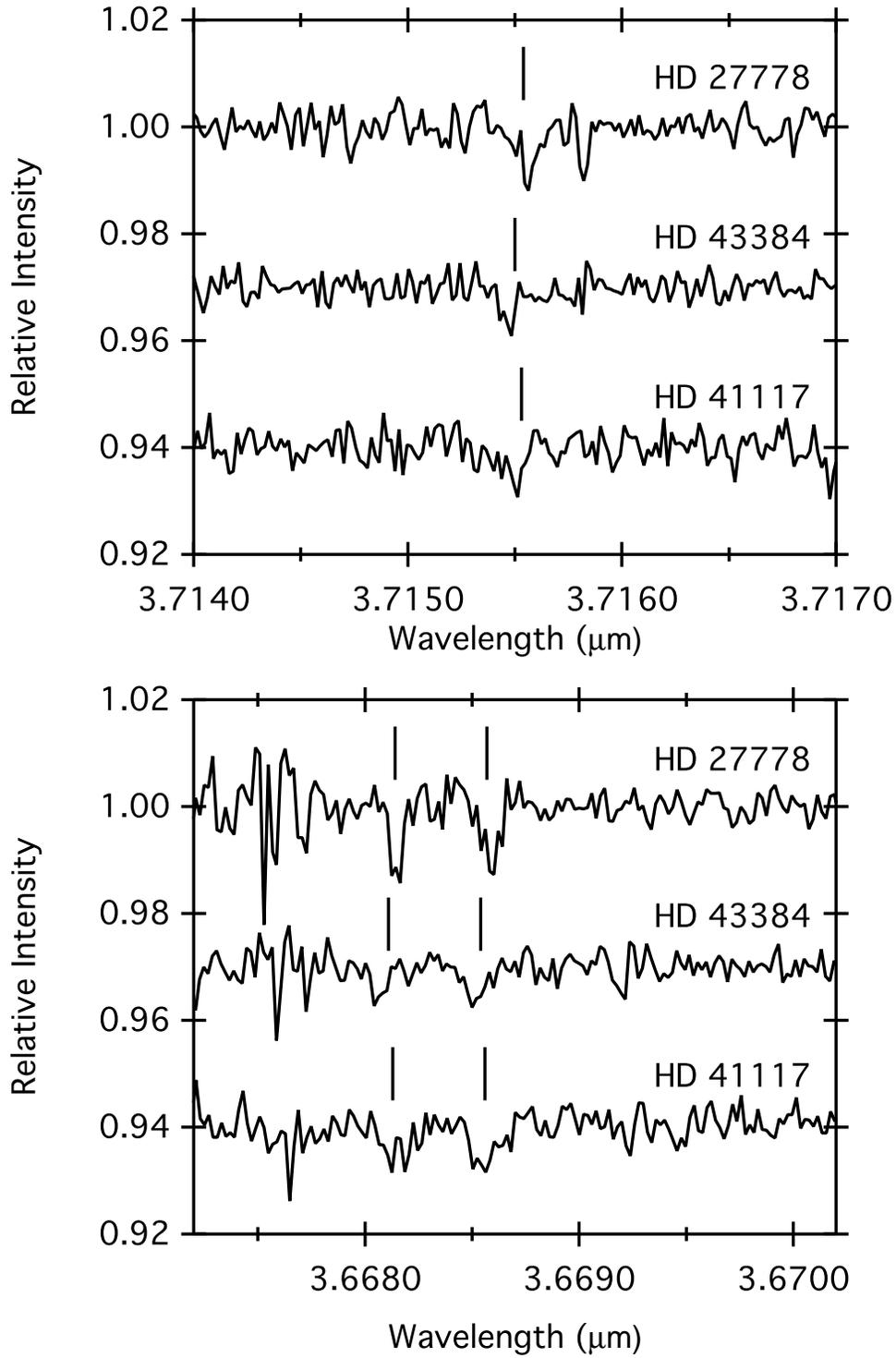}
\caption{Spectra of HD~27778, HD~43384, and HD~41117 showing absorption due to the $R(1,1)^{u}$ and $R(1,0)$ (bottom) and $R(1,1)^{l}$ (top) transitions of H$_3^+$. Vertical lines above spectra mark the expected positions of absorption features given previously determined interstellar gas velocities.
}
\label{fig_astro}
\end{figure}

Raw data images were processed using the CRIRES pipeline version 2.2.1. Standard calibration techniques, including subtraction of dark frames, division by flat fields, interpolation over bad pixels, and correction for detector non-linearity effects, were utilized. Consecutive A and B nod position images were subtracted from each other to remove sky emission features. One-dimensional spectra were extracted from these images using the {\it apall} routine in \textsc{iraf}\footnote{http://iraf.noao.edu/}, and any remaining bad pixels were interpolated over. Spectra were then imported to Igor Pro\footnote{http://www.wavemetrics.com/}, and all spectra from each nod position were added together. Wavelength calibration of the summed A and summed B spectra was performed using atmospheric absorption lines (accurate to $\pm1$~km~s$^{-1}$), after which the spectra from the A and B nod positions were averaged onto a common wavelength scale.\footnote{Combination of A and B spectra is done after wavelength calibration due to a slight (about one-half pixel) shift in wavelength along the dispersion direction between the two nod positions.} In order to remove atmospheric absorption features, science target spectra were divided by spectra of telluric standard stars. Resulting spectra are shown in Figure \ref{fig_astro}.

Absorption features due to H$_3^+$ were fit with Gaussian functions for the purpose of determining equivalent widths, velocity FWHM, and interstellar gas velocities. Equivalent widths were used to compute column densities in the lower state, and the \textit{para}-H$_3^+$ column density was then determined by taking a variance-weighted average of the values found from the $R(1,1)^u$ and $R(1,1)^l$ transitions. As we in this paper will discuss our results in terms of the \textit{para}$-$to$-$total ratio (\textit{para} fraction), we here define these parameters for H\dtwo~and H\dthree\jon~as:
\begin{eqnarray}
p_2 = [para-{\rm H}_2]/[{\rm H}_2]		\nonumber\\
p_3 = [para-{\rm H}_3^+]/[{\rm H}_3^+]	\nonumber
\end{eqnarray}
All of the absorption line parameters and derived column densities are presented in Table \ref{tab:obs}, along with the inferred $p_3$ and excitation temperatures, $T({\rm H_3^+})$. Also shown are H$_2$ column densities, H$_2$ {\it para} fractions ($p_2$), and $T_{01}$ presented in \citet{2002ApJ...577..221R, 2009ApJS..180..125R}. 

{The three new detections of H$_3^+$ presented in this paper increase the total number of diffuse cloud sight lines with measurements of column densities of the \textit{ortho} and \textit{para} forms of both H$_3^+$ and H$_2$ from six to nine. Two of these sight lines---HD 27778 and HD 41117\footnote{Note that these sight lines are frequently referred to by the alternate identifiers 62~Tau and $\chi^2$~Ori, respectively.}---were observed previously but H$_3^+$ was not detected \citep{2002ApJ...567..391M, 2012ApJ...745...91I}.  Column densities in the (1,0) and (1,1) levels reported in this paper are consistent with the previously reported upper limits. The newly reported values of $p_3$ and $p_2$ follow the same trend shown by the six older data points in \citet{2011ApJ...729...15C} and \citet{Crabtree13112012}.  This continues to demonstrate that in diffuse clouds the excitation temperatures of H$_2$ and H$_3^+$ do not agree with each other, even when both species are observed in the same line of sight. This developing trend, and the lack of an adequate explanation from simple chemical models, is part of the reason for the present study.}

\begin{deluxetable}{llccc}
\tablewidth{0pt}
\tabletypesize{\small}
\tablecaption{Absorption line parameters and derived values \label{tab:obs}}
\tablehead{ & & \colhead{HD 27778\tablenotemark{a}} & \colhead{HD 41117\tablenotemark{b}} & \colhead{HD 43384\tablenotemark{b}}}
\startdata
$v_{\rm LSR}$~$R(1,1)^u$ & (km~s$^{-1}$) & 5.2 & 5.7 & $-$1.5 \\
$v_{\rm LSR}$~$R(1,0)$ & (km~s$^{-1}$) & 5.9 & 2.5 & $-$0.1 \\
$v_{\rm LSR}$~$R(1,1)^l$ & (km~s$^{-1}$) & 6.9 & 2.2 & $-$0.8 \\[2pt]
\hline\\[-6pt]
$FWHM$~$R(1,1)^u$ & (km~s$^{-1}$) & 4.7 & 8.4 & 5.6 \\
$FWHM$~$R(1,0)$ & (km~s$^{-1}$) & 6.5 & 7.4 & 8.5 \\
$FWHM$~$R(1,1)^l$ & (km~s$^{-1}$) & 7.3 & 4.9 & 5.4 \\[2pt]
\hline\\[-6pt]
$W_{\lambda}$~$R(1,1)^u$ & (10$^{-6}$~$\mu$m) & 0.93$\pm$0.14 & 1.11$\pm$0.23 & 0.49$\pm$0.11 \\
$W_{\lambda}$~$R(1,0)$ & (10$^{-6}$~$\mu$m) & 1.00$\pm$0.17 & 0.82$\pm$0.20 & 0.73$\pm$0.15 \\
$W_{\lambda}$~$R(1,1)^l$ & (10$^{-6}$~$\mu$m) & 0.91$\pm$0.21 & 0.56$\pm$0.15 & 0.55$\pm$0.13 \\[2pt]
\hline\\[-6pt]
$N(J,K)$~$R(1,1)^u$ & (10$^{13}$~cm$^{-2}$) & 3.87$\pm$0.58 & 4.60$\pm$0.97 & 2.03$\pm$0.47 \\
$N(J,K)$~$R(1,0)$ & (10$^{13}$~cm$^{-2}$) & 2.53$\pm$0.42 & 2.07$\pm$0.52 & 1.85$\pm$0.37 \\
$N(J,K)$~$R(1,1)^l$ & (10$^{13}$~cm$^{-2}$) & 4.18$\pm$0.94 & 2.55$\pm$0.68 & 2.54$\pm$0.62 \\[2pt]
\hline\\[-6pt]
$N (1,1)$ $-~$H$_{3}^{+}$ & (10$^{13}$~cm$^{-2}$) & 3.96$\pm$0.22 & 3.22$\pm$1.45 & 2.22$\pm$0.36 \\
$N (1,0)$ $-~$H$_{3}^{+}$ & (10$^{13}$~cm$^{-2}$) & 2.53$\pm$0.42 & 2.07$\pm$0.52 & 1.85$\pm$0.37 \\
$p_3$ & & 0.61$\pm$0.04 & 0.61$\pm$0.12 & 0.55$\pm$0.06 \\
$T({\rm H}_3^+)$ & (K) & 29$\pm$4 & 29$\pm$13 & 38$\pm$11 \\[2pt]
\hline\\[-6pt]
$\log[N(0)]$ $-~$H$_2$ & & 20.64$\pm$0.05 & 20.51$\pm$0.10 & 20.59$\pm$0.10 \\
$\log[N(1)]$ $-~$H$_2$ & & 20.27$\pm$0.10 & 20.22$\pm$0.10 & 20.54$\pm$0.18 \\
$p_2$ & & 0.70$\pm$0.05 & 0.66$\pm$0.07 & 0.53$\pm$0.12 \\
$T_{01}$ & (K) & 56$\pm$5 & 60$\pm$7 & 74$\pm$15
\enddata
\footnotetext{H$_{2}$ data collected from (a) \citet{2002ApJ...577..221R} and (b) \citet{2009ApJS..180..125R}.}
\end{deluxetable}

\section{Model}\label{sec:model}
\subsection{Physical model}
We have utilized the gas-grain chemical model ``ALCHEMIC'' developed by \citet{2010A&A...522A..42S}, where a detailed description of the code and its performance is presented. {Below we give a brief explanation of the code. In these low-density environments, surface interaction is not expected to play an important role in the chemistry. Never the less this is included in our model and explained in detail in \citet{2010A&A...522A..42S}. }

{To calculate UV ionization and dissociation rates, several tens of photoreaction rates are updated using the calculations of \citet{2006FaDi..133..231V}. A UV photodesorption yield for surface species of 10$^{-3}$ is assumed \citep[see e.g.][]{2009A&A...496..281O, 2009ApJ...693.1209O}. The self-shielding of H$_2$ from photodissociation is calculated by Eq. (37) from \citet{1996ApJ...468..269D}, {assuming a total extinction of 0.5 mag with $N($H$) \approx 1.6 \times 10^{21}$ cm$^{-2}$, resulting in a $N($H$_2) \approx 4\times 10^{20}$ cm$^{-2}$.} The shielding of CO by dust grains, H$_2$, and its self-shielding is calculated using {the} precomputed table of \citet[][see Table 11]{1996A&A...311..690L}. }

%{The gas-grain interactions include sticking of neutral species and electrons to uniformly-sized dust grains with 100\% probability, release of frozen molecules by thermal, CRP-, and UV-induced desorption, dissociative recombination and radiative neutralization of ions on charged grains, and grain re-charging. 
%We do not allow H$_2$ to stick to grains because the binding energy of H$_2$ to pure H$_2$ mantle is low, $\sim$ 100 K \citep{1972NPhS..237...99L}, and it freezes out in substantial quantities only at temperatures below $\approx$ 4 K. Probability of reactive desorption upon a surface recombination is considered to be 1\%. 
%We assume that each 0.1$\mu$m spherical olivine grain provides $\approx 2\times 10^6$ surface sites, and that surface recombination proceeds solely through the Langmuir-Hinshelwood formation mechanism. Following interpretations of experimental results on the formation of molecular hydrogen on dust grains \citep{1999ApJ...522..305K}, we employ the standard rate equation approach to the surface chemistry without H and H$_{2}$ tunneling either through the potential walls of the surface sites or through reaction barriers.} 

In this work we are primarily concerned with reproducing observed \textit{ortho}$-$ and \textit{para}$-$abundances of H\dthree\jon~and H\dtwo. We conduct a parametric study using our time-dependent chemical model. For this goal, gas density, DR and CR ionization rates, as well as initial \textit{ortho}/\textit{para} ratio of H$_2$ and chemical age, are varied within the ranges typical of the diffuse ISM, and the modeled \pHtva~and \pHtre~values are compared with the observed values. 
{There are still large uncertainties as to the initial abundances in diffuse clouds, concerning especially the depletion factor of metals and the degree of ionization \citep[see e.g.][]{1995A&A...297..251L, 2006ARA&A..44..367S, 2009ApJ...700.1299J}. To study the impact of different initial abundances we implement} both the ``low metal'' {and ``high metal''} abundances from \citet{1982ApJS...48..321G} and \citet{1998A&A...334.1047L}. Furthermore, we consider the neutral and fully ionized cases of initial abundances. 
Diffuse molecular clouds have typical densities of 10 - 100 cm$^{-3}$ with typical extinction of $A_{V} = 0.5$~mag \citep{2006ARA&A..44..367S}. 
{The CR ionization rate in the diffuse ISM is a matter of debate \citep[see e.g.][]{2007ApJ...671.1736I, 2009ApJ...694..257I, 2012A&A...537A...7R, 2012ApJ...745...91I}. It is clear that the CR ionization rate plays a crucial role in the chemical evolution owing to its importance in the formation of H$_3^+$ through the ionization of H$_2$ and it also has a significant effect on the \textit{para}-fractions of H$_2$ and H$_3^+$. Therefore we consider a wide range of CR ionization rates between $10^{-17} - 10^{-15}$ s$^{-1}$.
}
The lifetime of giant molecular clouds is typically in the range of several $10^{7}$ years \citep{2007ARA&A..45..565M}, giving us an upper limit for the time scales of diffuse cloud evolution. Our cloud models are calculated within a time span of 10$^{6}$ years, but we also investigate the time effects on results for longer time spans of 10$^7$ years. As a ``Standard'' model (``S'') we consider a hydrogen gas density $n_{\rm H}$~=~10 cm$^{-3}$, $\zeta_{\rm CR}$ = $10^{-16}$ s$^{-1},~A_{\rm V}$ = 0.5 mag, and temperatures $10 -100$~K, adopting the H$_{3}^{+}$ DR rates of \citet{2004PhRvA..70e2716M}. The initial \textit{ortho}/\textit{para} ratio of H$_2$ is {still unknown for the ISM, therefore we investigate two values: 1:10, characteristic of low temperatures, and 3:1, the value characteristic for warm temperatures $\gtrsim 100$ K \citep[see e.g.][]{1999ApJ...516..371S, 2011ApJ...739L..35P}. }

\subsection{Chemical network}
We used a smaller version of the chemical network developed by \citet{2013ApJS..207...27A}, {which is based on the osu.2007 ratefile with recent updates to reaction rates, which we have modified such as only H-bearing reactions with $<4$ H atoms, $<4$ C atoms and molecules made of $<8$ atoms are cloned}. In this study we extended it to include the \textit{ortho}$-$\textit{para} states of H$_{2}$, H$_{2}^{+}$ and H$_{3}^{+}$, and related nuclear spin-state exchange processes. Reaction rates for a small number of reactions have already been measured or theoretically predicted. For this, we have added rates from several sources \citep{1990JChPh..92.2377G, 2004A&A...418.1035W, 2004A&A...427..887F, 2009A&A...494..623P, 2011PhRvL.107b3201H}, including reaction rates for the H$_{3}^{+}$ + H$_{2}$ system by \citet{2009JChPh.130p4302H}. For any remaining reactions involving H$_{2}$, H$_{2}^{+}$, or H$_{3}^{+}$ with unknown rates, we extracted reactions containing these species and employed a separation scheme, similar to that from \citet{2013A&A...554A..92S}, in order to generate \textit{ortho}$-$ and \textit{para}$-$variations of the reactions. We illustrate the process of our separation scheme with the following example:
\begin{eqnarray}
\small
{\rm H}_3^+ + {\rm CO} \rightarrow {\rm HCO}^{+} + {\rm H}_{2} \left\{
\begin{array}{l@{~~~}l}
\textit{ortho}-{\rm H}_{3}^{+} + {\rm CO} \rightarrow {\rm HCO}^{+} + \textit{ortho}-{\rm H}_{2} \\[1mm]
\textit{para}-{\rm H}_{3}^{+} + {\rm CO} \rightarrow {\rm HCO}^{+} + \textit{para}-{\rm H}_{2} \\[1mm] 
\textit{para}-{\rm H}_{3}^{+} + {\rm CO} \rightarrow {\rm HCO}^{+} + \textit{ortho}-{\rm H}_{2} \nonumber
\end{array}
\right.
\normalsize 
\end{eqnarray}
where the left side is the original reaction, and the right shows the pathways resulting from the separation scheme. The reaction rate $R$ is the same as the original rate $R_{\rm org}$ for the first pathway, while the reaction for \textit{para}-H$_{3}^{+}$ has two possible sets of products, and the reaction rate is divided between the two pathways, $R_{\rm org}/2$. This branching ratio is due to spin statistics \citep[see e.g.][]{Crabtree2013, OkaFestschrift}. Contrary to \citet{2013A&A...554A..92S}, we allow reactions without H\dthree\jon~or H\dtwo\jon~as reactants to form not only \textit{para}-H$_2$, but also \textit{ortho}$-$H\dtwo~with an energy barrier of $\gamma=170$~K. The final network consists of ~1\,300 species connected by ~40\,000 reactions. This networks includes gas-grain interactions and surface reactions. We assume that H$_2$ is formed with a 3:1 \textit{ortho}:\textit{para} ratio, aligned with what is generally assumed in models and also supported by the first confirmation from laboratory experiments \citep{2010ApJ...714L.233W}. However, the \textit{ortho}$-$\textit{para} interconversion on dust surfaces are still highly uncertain \citep{2011PCCP...13.2172C}. {In Table~\ref{tab:models} we summarize the different models used to study the effects of physical and chemical parameters. }

\begin{deluxetable}{llllll}
\centering
\tablewidth{0pt}
\tabletypesize{\scriptsize}
\tablecaption{Summary of studied models. \label{tab:models}}
\tablehead{ 
\colhead{Model}	&	\colhead{$n_{H}$}	&	\colhead{CR rate}	&	\colhead{Time}		&	\colhead{DR rates}					&	\colhead{Inital}				\\
			&	\colhead{[cm$^{-3}$]	}&	\colhead{[s$^{-1}$]}	&	\colhead{[years]}	&	\colhead{[reference]}					&	\colhead{H$_2$ o:p}			}
\startdata
S 			&	10				&	10$^{-16}$			&	10$^{6}$			&	\citet{2004PhRvA..70e2716M}			&	1:10						\\
N100 		&	100				&	10$^{-16}$			&	10$^{6}$			&	\citet{2004PhRvA..70e2716M}			&	1:10						\\
N1000		&	1000				&	10$^{-16}$			&	10$^{6}$			&	\citet{2004PhRvA..70e2716M}			&	1:10						\\
C15 			&	10				&	10$^{-15}$			&	10$^{6}$			&	\citet{2004PhRvA..70e2716M}			&	1:10						\\
C17 			&	10				&	10$^{-17}$			&	10$^{6}$			&	\citet{2004PhRvA..70e2716M}			&	1:10						\\
T 			&	10				&	10$^{-16}$			&	10$^{7}$			&	\citet{2004PhRvA..70e2716M}			&	1:10						\\
D 			&	10				&	10$^{-16}$			&	10$^{6}$			&	\citet{santos:124309}					&	1:10						\\
O			&	10				&	10$^{-16}$			&	10$^{6}$			&	\citet{2004PhRvA..70e2716M}			&	3:1						\\
2X			&	10				&	10$^{-16}$			&	10$^{6}$			&	2$\times$	\citet{2004PhRvA..70e2716M}	&	1:10						
%4X			&	10				&	10$^{-16}$			&	10$^{6}$			&	4$\times$	\citet{2004PhRvA..70e2716M}	&	1:10	
\enddata
\end{deluxetable}

\subsubsection{H$_{3}^{+}$ dissociative recombination}\label{sec:H3dr}
One of the crucial reactions for the hydrogen chemistry in the diffuse interstellar gas is the DR of H$_3^+$ with free electrons:
\begin{eqnarray}
{\rm H}_3^+ + e^- \rightarrow \left\{
\begin{array}{l@{~~~}l}
{\rm H} + {\rm H} + {\rm H} \\[1mm]
{\rm H}_2 +{\rm H} 
\end{array}
\right.\nonumber
\label{eq:dr}
\end{eqnarray}
The DR process is one of the dominant destruction processes for H$_3^+$ in the environments considered here, it influences the ionization balance by removing the reactive H$_3^+$ ion and yielding neutral fragments. Owing to its astrophysical relevance, the DR of H$_3^+$ is a much-studied process, both experimentally and theoretically. The outcome of H$_3^+$ DR studies has varied over the years, and to date there are more than 30 published experimental rate coefficients for this reaction. Historically, there were orders of magnitude discrepancies between different experimental approaches at times, with flowing and stationary afterglow experiments resulting in substantially lower rates than storage ring experiments, which have become the prevalent method for DR studies since the early nineties. However, the afterglow experiments have been re-evaluated recently \citep{glosik09}, and the inclusion of ternary collisions in the analysis has led to a good overall agreement between different methods concerning the absolute scale of the low-energy rate coefficient for H$_3^+$. A review on the early measurements and disagreements can be found in \citet{larsson00}; here we will focus on the current state-of-the-art and the best rate coefficient to use for our purpose. 

The commonly accepted experimental value for the absolute DR rate is given as a thermal rate coefficient in \citet{2004PhRvA..70e2716M}, as a function of temperature:
\begin{equation}
r_{\rm DR, McCall}=-1.3\times 10^{-8} + 1.27\times 10^{-6} T^{-0.48}\label{eq:DRmccall}
\end{equation}
The underlying measurements were carried out using a supersonic expansion ion source at the CRYRING storage ring and confirmed over a wide range of relative energies by subsequent measurements at the Test Storage Ring (TSR) in Heidelberg \citep{kreckel05,kreckel10}. However, the latest of these studies \citep{kreckel10, petrignani11} revealed that the expansion ion source delivered much hotter ions than previously assumed, and the best characterized measurements were carried out at $\sim$370\,K \citep{petrignani11}. The implications for the rate coefficient at $T<100$\,K are uncertain. 

Theoretically, the DR of H$_3^+$ has proven difficult to describe. The classical picture of direct DR that proceeds through a curve-crossing of a dissociative state does not apply for H$_3^+$, and thus initial studies predicted a very low rate coefficient. This picture has changed when more modern studies in full-dimensionality became available and identified the Jahn-Teller effect as the driving force behind H$_3^+$ DR \citep{kokoouline01, kokoouline03, kokoouline03b}. With these major advances, the more recent calculations of \citet{santos:124309} agree quite well with the storage ring measurements for the absolute recombination rate, while discrepancies remain for the detailed energy-dependence \citep{petrignani11}. 
%Closer scrutiny reveals that even between the thermal rate coefficients given in \citet{2004PhRvA..70e2716M} and \citet{santos:124309} a difference of almost a factor of 2 remains at low temperature. At present we accept this range as the uncertainty in the absolute rate coefficient of H$_3^+$ DR and we will test the impact that the variation in this important parameter has on our model by running simulations with either value. 
{We adopt rate coefficients converted to the Kooji formulation, by adjusting the coefficients for best agreement in our temperature range of 10$-$100 K:
}
\begin{equation}
r_{\rm DR, dos Santos}=-1.1\times 10^{-7} \times (T/300)^{-0.52} \label{eq:DRsantos}
\end{equation}
{Deviations from the original fit in this temperature range are below $\sim$ 20 \%}. 

The situation gets even more complicated when one considers the dependence of the DR rate on the nuclear spin of H$_3^+$. The calculations of \citet{santos:124309} predict that at $T<100$\,K the DR rate is dominated by the \textit{para}-H$_3^+$ states and at 10\,K the rate for $(1,0)-$H$_3^+$ is more than an order of magnitude lower than for $(1,1)-$H$_3^+$. These theoretical values are supported by recent afterglow results that see a similar trend \citep{2011PhRvL.106t3201V}. It should be kept in mind, though, that the calculated rate coefficient at these temperatures depends on the precise position of Rydberg resonances that are difficult to predict, and the initial calculations actually showed the opposite trend \citep{kokoouline03b}. 

The first measurements that showed a nuclear spin dependence of the H$_3^+$ DR rate were carried out at the TSR \citep{kreckel05}, however, only a slight enhancement of the rate for \textit{para}$-$H$_3^+$ was seen. The same trend has been observed in more recent studies by \citet{tom09} and \citet{kreckel10}. The caveat of these measurements is again the fact that the ions were probably not in their rotational ground states. More detailed studies with state-selected molecular ions are clearly desirable for this important reaction. For now, we implement both extremes in our model calculations, the almost exclusive dominance of the low-energy DR by \textit{para}-H$_3^+$, as predicted by the calculations of \citet{santos:124309}, and basically equal rate coefficients for \textit{ortho}$-$ and \textit{para}$-$H$_3^+$, as seen in the storage ring measurements. 

\subsection{Dominant pathways}
First, to ease our detailed analysis of the chemical processes relevant for the evolution of H$_2$ and H$_3^+$ spin states, we identify the dominant pathways in their chemistry. For that we reduce our initially huge gas-grain chemical network with surface reaction to a much smaller set of reactions using our ``Automatic Reduction Technique'' (ART) tool \citep[see, e.g.,][]{2006ApJ...647L..57S}. The reduced network consists of only 20 species and 144 reactions, and is accurate within $<$~10\% in the studied parameter space for the abundances of H$_2$ and H$_3^+$ (see Table~2).

The dominant pathways include ionization by CR particles and UV photons as well as DR and ion-molecule reactions. In comparison to \citet{2011ApJ...729...15C}, who considered a much simpler chemistry, we find additional pathways that, when combined, bear a significant effect on the resulting \textit{para}-fraction values. Because of the importance that H\jon~plays for the H\dtwo~\textit{ortho}$-$\textit{para} ratios, it becomes necessary to consider the photodissocation of H\dtwo\jon, a major formation pathway of H\jon. Equally important are the DR of the simple molecular ions made of H, C, O, which affects the rate of the H\dthree\jon~DR reaction. We also find that, {with the exception being S}, elements heavier than oxygen do not have any impact on the hydrogen chemistry. {We note that S$^{+}$ does not bear any large significance in comparison to C$^{+}$ for determining the time-dependent \textit{ortho}$-$\textit{para} abundances of H$_{2}$ and H$_{3}^{+}$ and the effect on the total degree of ionization is at most $\sim$10\%. }

Much as \citet{2011ApJ...729...15C}, we find that surface processes do not play a significant role in the chemistry of H$_2$ and H$_3^+$. This is because we begin our chemical simulations with hydrogen already being essentially in molecular form {and the densities being too low for any efficient surface chemistry to take place. } 
The gas-phase formation of H\dtwo~proceeds via slow neutral-neutral collisions of H atoms or ion-molecule reactions like H\dthree\jon~+ O $\rightarrow$ OH\jon~+ H\dtwo, and is not efficient. 

Aside from DR reactions, there is a small number of ion-molecule reactions that aid in the removal of H\dthree\jon. The two most notable of the less effective destruction pathways is H\dthree\jon~reacting with OH or O, forming H$_{2}$O$^{+}$ or OH$^{+}$, respectively, and H$_{2}$. As we show below, at gas hydrogen density of $\sim$ 10 cm$^{-3}$ these reactions can slow down the increase in \pHtre~and causing small changes to occur with time. 

Furthermore, we found inconsistencies in the main public astrochemical networks related to the reaction rates of the photodissociation of H\dthree\jon. Because H\dthree\jon~is expected to be in the ground state in the diffuse interstellar medium, a photon with an energy of almost 20 eV is required for it to be photo-dissociated \citep{1988JChPh..89.2235T}. This essentially means that H\dthree\jon~photodissociation is inactive under ISM conditions. Theoretical modeling agrees with this \citep[see e.g.][]{1987IAUS..120...51V, 2006FaDi..133..231V}. In the UMIST database \citep[latest version UMIST 2012;][]{2013A&A...550A..36M}, the correct photodissociation rates for H\dthree\jon~are incorporated. However, this rate is seven orders of magnitude higher in other databases, such as KIDA\footnote{\url{http://kida.obs.u-bordeaux1.fr/}} and OSU\footnote{\url{http://www.physics.ohio-state.edu/~eric/research.html}}. Because our network is based on the recent osu.2009 network, it had an incorrect rate for the H$_3^+$ photodissociation on the order of $\sim 10^{-8}$ s$^{-1}$, which we have corrected in this study by adopting the corrected value of $5\times 10^{-15}$ s$^{-1}$ from \citet{1987IAUS..120...51V}. 

\section{Results}\label{sec:results}
\subsection{Parameter effects}
Implementation of the chemical processes as well as properties of the environment, including density, temperature, and the choice of the chemical age, are the major actors for chemical kinetics modeling. Among these factors, we find that the adopted values of the DR rates have the greatest effect on modeled abundances of \textit{ortho}$-$ and \textit{para}$-$H$_3^+$, while the chemical age, the density, and the CR ionization rate are less important. In our models the \textit{ortho}$-$\textit{para} chemistry for H$_2$ is already at a quasi steady-state after $\sim 10^{5}$ years, which is shorter than our standard adopted chemical age of $\sim 10^{6}$ years. Hence, unless otherwise specified, we discuss only the time-dependent evolution of the H$_3^+$ \textit{ortho}$-$\textit{para} chemistry. 
{
We calculated the "S" model using two initial abundances with "high" and "low" metal abundances, and found that there is a $< 10\%$ difference in abundances of key molecules. Considering an initially ionized medium (except for H, He, O, N) only shortens the time needed for \textit{para}-fractions to reach their equilibrium values. More importantly, in both these cases, $p_2$ and $p_3$ remain unaffected and hence our conclusions are not affected by the choice of initial abundances beyond times $\gtrsim 10^6$ years. 
}

\begin{figure*}[!p]
\centering
\includegraphics[width=0.65\textwidth]{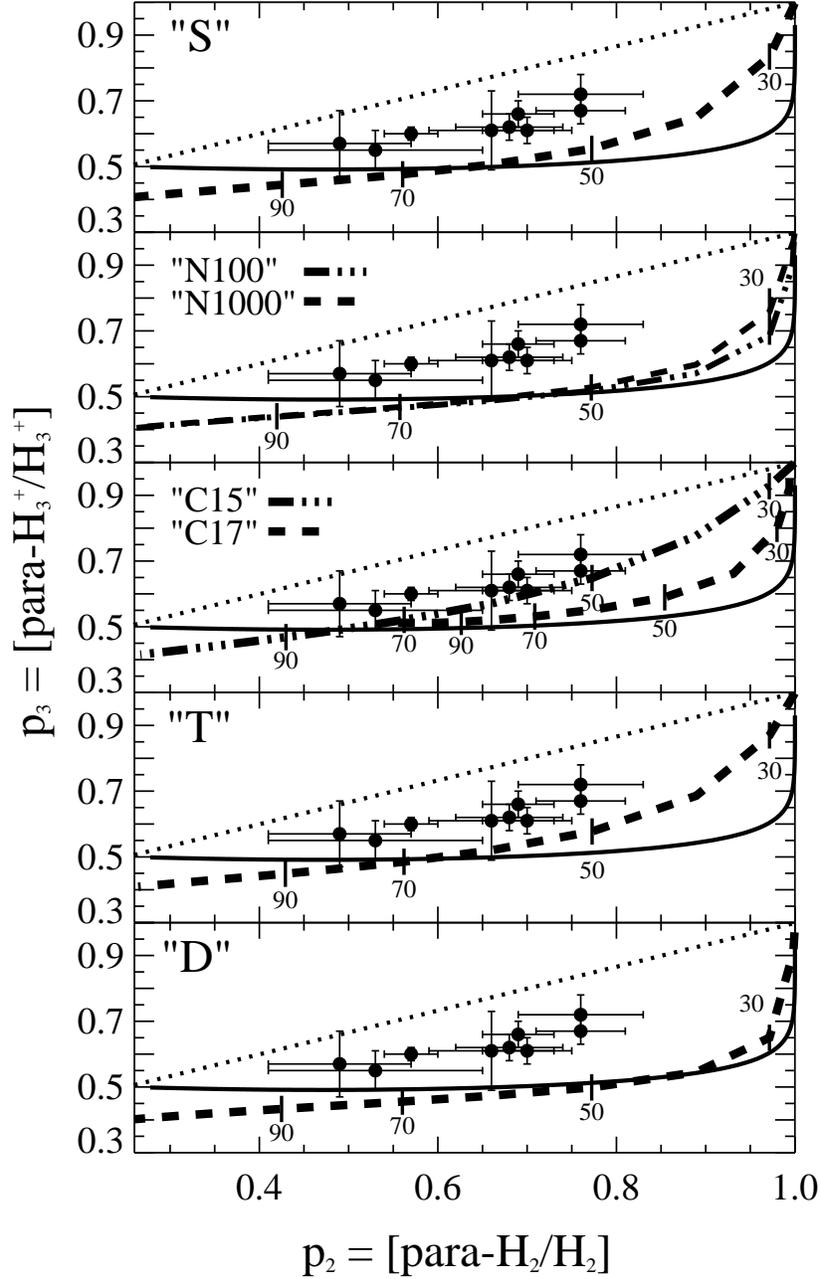} 
\caption{Comparison of the effects on \pHtre~and \pHtva~values from density, CR ionization rate, time and DR reaction rates compared between our different models, calculated for kinetic temperatures 10 - 100 K. The dotted line is the nascent distribution and solid line the thermal distribution. The calculated $p_{2}$ and $p_{3}$ values at temperatures 30, 50, 70 and 90 K are marked specifically in the figures. Observation are shown with 1$\sigma$ error bars. {The "S`` model is the standard model with $n_H = 10$ cm$^{-3}$, $\zeta_{CR} = 10^{-16}$ s$^{-1}$, A$_{\rm V}$ = 0.5 mag, adopting the H$_{3}^{+}$ DR rates of \citet{2004PhRvA..70e2716M} and calculated for temperatures  $10 -100$~K. Compared to the "S`` model, in the "N100`` and "N1000`` models we have increased the hydrogen gas density to $n_H = 100$ and $1000$ cm$^{-3}$ respectively, in the "C15`` and "C17`` models the CR ionization are set to $\zeta_{CR} = 10^{-15}$ and $10^{-17}$ s$^{-1}$ respectively, in the "T`` model the time scale is increased to $10^7$ years and in the "D`` model the H$_3^+$ DR rate is set equal to that predicted by \citet{santos:124309}.}
}
\label{fig:OPR}
\end{figure*} 

In Figure~\ref{fig:OPR} we show the resulting \pHtre~and \pHtva~values for our models with varying density {(models ``N100'' and ``N1000'')}, CR ionization rate {(models ``C17'' and ``C15'')}, time (model ``T'') and reaction rates for the H$_{3}^{+}$ DR reactions (model ``D''), compared to the "Standard`` model (model ``S''). 

The ``S'' model does not agree with the trend observed towards diffuse clouds (see Figure~\ref{fig:OPR}), but is very similar to the distribution calculated by \citet{2011ApJ...729...15C} (see their Figure 8), where they considered only the H$_{3}^{+}$ formation and its destruction by DR. Our models are also compared to thermalized distributions, calculated using energy levels from \citet{2001JMoSp.210...60L}. The nascent distribution, where \pHtre~= 1/3 + 2/3\pHtva~ is also plotted. For this distribution we assume that H\dthree\jon~is exclusively formed from reaction H\dtwo\jon~+ H\dtwo~and we use the branching ratios adopted from nuclear spin selection rules of \citet{1977MolPh..34..477Q} and \citet{2004JMoSp.228..635O}. Furthermore, we assume that the CR ionization of H\dtwo~does not affect the H\dtwo\jon~nuclear spin configuration.

{
We consider three CR ionization rates, $10^{-17}, 10^{-16}$ and $10^{-15}$ s$^{-1}$ (models ``S'', ``C17'' and ``C15'' respectivelly). The ``C17'' model decreases the $p_3$ values while increasing $p_2$ values, worsening the agreement compared to the ``S'' model. Meanwhile, the ``C15'' model significantly improves the fit such that predicted \textit{para}-fractions go through the observed data points. Variations in the CR ionization rate affects the ionization degree, affecting the abundances and \textit{para}-fractions of ions. 
}

Although the diffuse interstellar clouds are likely clumpy, they are often assumed to be homogeneous in astrochemical models. We consider {three} gas densities typical of diffuse ISM, 10, 100 and 1000 cm$^{-3}$ (model ``S'', ``N100'' and ``N1000'' respectivelly). We find that results for model ``S'' with a hydrogen density of 10 cm$^{-3}$ are closer to the observed values of $p_{3}$. The reason for the higher \textit{para}-H$_3^+$ fraction with lower density is {again} the ionization degree of the medium, which is largely determined by H$^+$ and C$^+$. At lower densities the neutralization of the medium through DR reactions proceeds slower, causing a higher ionization degree, however it is not enough to affect the \pHtva~value. 

While we can not predict column densities as our models have no spatial dimension, we can compare the ratios of calculated H$_{3}^{+}$\,/\,H$_{2}$ abundances to the observed column densities, which should agree unless extreme clumpiness occurs. Doing this, for the final abundances at 1 Myr, our models predicts for temperatures 30 $-$ 70 K abundances ratios H$_{3}^{+}$\,/\,H$_{2}$ = $2.13 - 8.70 \times 10^{-7}$ for 10 cm$^{-3}$ {(model ``S''), $1.12 - 2.34 \times 10^{-7}$ for 100 cm$^{-3}$ {(model ``N100'') and $2.37 - 4.77 \times 10^{-8}$ for 1000 cm$^{-3}$} (model ``N1000'')}, while observed column density ratios are typically $\sim 10^{-7}$ \citep[][and this study]{2011ApJ...729...15C, 2012ApJ...745...91I}. This means that modeled abundances for the gas density 100 cm$^{-3}$ agree better with the observed H$_{3}^{+}$\,/\,H$_{2}$ ratios, while the p$_3$ values still disagree. It also becomes clear that the temperature has a larger effect on abundances at 10 cm$^{-3}$ than at 100 cm$^{-3}$, considering the predicted wide range of abundances for the 10 cm$^{-3}$ model.
{
The effects of the CR ionization modeled in the ``C17'' and ``C15'' models also affects the  H$_{3}^{+}$\,/\,H$_{2}$ abundances ratios, and for the former we calculate H$_{3}^{+}$\,/\,H$_{2}$ = $1.30 - 2.10 \times 10^{-7}$ and for the latter H$_{3}^{+}$\,/\,H$_{2}$ = $6.25 - 7.75 \times 10^{-7}$. This means that the ``C15'' model improves the agreement in the \textit{para}-fraction distributions and the calculated H$_3^+/$H$_2$ abundance ratios are within reasonable agreement to observed column densities. 
}

\subsubsection{Time-dependence}
\begin{figure}[!htb]
\centering
\includegraphics[width=0.5\textwidth, angle=90]{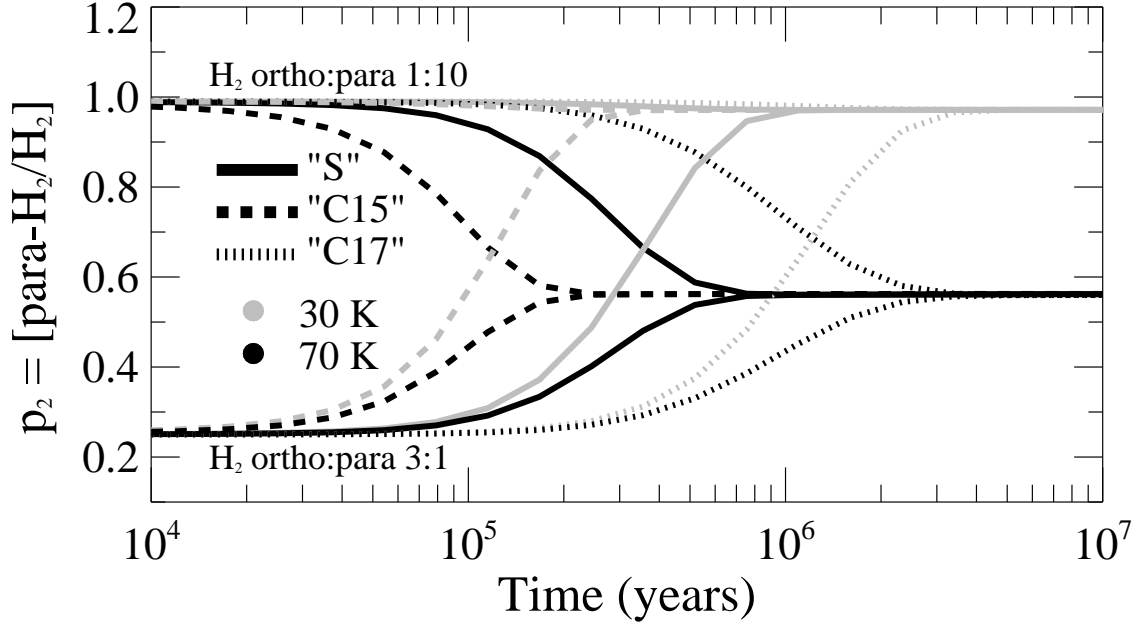} 
\caption{Evolution of $p_{2}$ with time. Results from the ``S'' model are shown by the solid line, the ``C17'' model by dashed line {and the ``C15'' model by dot-dashed line}. Two different initial H$_2$ \textit{ortho}:\textit{para} are tested: 1:10 and 3:1. {H$_2$ para-fractions} at 30 K are shown in gray and in black 70 K. }
\label{fig:timeH2}
\end{figure} 
In Figure~\ref{fig:timeH2} we show the evolution of $p_{2}$ values for models ``S'', ``C15'' and ``C17'' as a function of time for two separate temperatures, 30 K and 70 K, and considering two initial H$_2$ \textit{ortho}/\textit{para} ratio, 3:1 (equilibrium ratio) and 1:10 (ratio of colder environments). 
At 30 K the $p_{2}$ values do not change significantly between models ``S'', ``C15'' and ``C17'', as they are close to the thermalized value already from the beginning. The ``S'' model reaches its 30 K steady-state value $p_2 \approx 0.97$ after $\sim 10^6$ years, while the 70 K value $p_2 \approx 0.56$ requires only $\sim 7\times 10^5$ years. However, in the ``C17'' model it takes longer, $\sim 3\times 10^6$ years at both 30 and 70 K, to reach the same steady-state values, {as opposed to $\sim 2\times 10^5$ years for the ``C15'' model}. This means that the results are strongly time-dependent, and the steady-state \pHtva~values have not changed, but merely the process of reaching the steady-state values is slowed down. The variation in CR ionization rate affects the production of H$^{+}$, the essential thermalization agent of H$_{2}$ and this further affects the time it takes to reach the thermalized \textit{ortho}/\textit{para} ratio. The calculated $para-$fractions become independent of the modeling assumptions on time scales $\ga 10^{7}-10^{8}$~years. 

\begin{figure}[!htb]
\centering
\includegraphics[width=0.5\textwidth, angle=90]{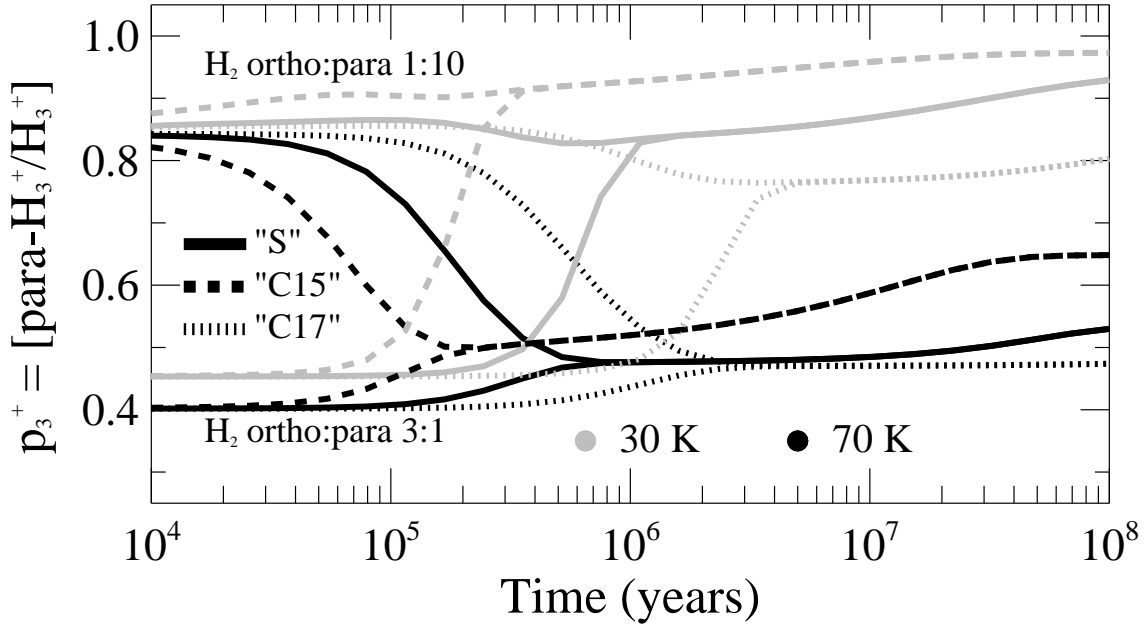} 
\caption{Evolution of $p_{2}$ with time. Results from the ``S'' model are shown by the solid line, the ``C17'' model by dashed line {and the ``C15'' model by dot-dashed line}. Two different initial H$_2$ \textit{ortho}:\textit{para} are tested: 1:10 and 3:1. {H$_2$ para-fractions} at 30 K are shown in gray and in black 70 K. }
\label{fig:timeH3}
\end{figure} 

Contrary to the H$_2$ evolution, the modeled \textit{ortho}$-$\textit{para} H$_3^+$ abundances have not reached steady-state by the chemical age of 10$^{6}$ years. Therefore, we study how \pHtre~evolves in the ``S'' model at later times up to 10$^{7}$ years (model ``T'', see Figure~\ref{fig:OPR}). In general, the \pHtre~values increase with temperature and the temperature distribution slowly approaches the nascent distribution. However, this process appears to require an unrealistic amount of time exceeding 10$^{8}$ {years} for models ``N100'' and ``N1000''. It is clear that even after only $\sim 10^{7}$ years the \pHtre~distribution is showing an appreciable difference to its value after 10$^6$ years.

This becomes much clearer in Figure~\ref{fig:timeH3} where the time evolution of $p_3$ is shown for the ``S'', ``C15'' and ``C17'' models up to 10$^8$ years. We can see that 10$^8$ years {are} necessary in order for the ``C15'' model to reach steady-state while the other models needs even longer time scales as the $p_3$ values are slowly increasing with time until they reach steady-state values of $p_3 \approx 0.97$ at 30 K and $p_3 \approx 0.67$ at 70 K. This steady increase in $p_3$ values is due to the slow redistribution of hydrogen from molecular into atomic form (H and H$^+$). We can conclude that it is clear that steady-state models are not appropriate in modeling the $p_3$ values of diffuse clouds. 

\subsubsection{Influence of DR rate coefficient}

The total rate and branching ratios for the DR of (1,0)$-$ and (1,1)$-$H\dthree\jon~greatly affects the modeled \pHtre~values (see Figure~\ref{fig:OPR}). In model ``D'' we compare the $p$-values of H\dthree\jon~and H\dtwo~for the chemical models adopting a branching ratio of $\sim $1:15 for the \textit{ortho}:\textit{para} H$_3^+$ DR rates, as predicted by \citet{santos:124309}. There are clear differences between the various models and the observed values, with the largest discrepancy arising at temperatures $\lesssim 60$ K, which are typical representative temperatures of the diffuse ISM \citep{2006ARA&A..44..367S}. The H$_3^+$ DR process with a preferred destruction of $(1,1)-$H\dthree\jon~results in \pHtre~values that are lower than the thermalized distribution at temperatures $> 30$ K. The \pHtre~values continue to decrease with time. As we adjust the DR branching ratio for (1,0)$-$ and (1,1)$-$H$_{3}^{+}$ towards unity, the overall \pHtre~values increase, most notably at low temperatures, and approximately at a ratio of $1:5$ the time dependence is reversed, and the \pHtre~values begin to slowly increase with time. 
%
% The key conclusion is that a DR branching ratio approaching unity for $(1,1)-$ and $(1,0)-$H$_3^+$ is necessary in order to reproduce the observed distributions. We note that the total DR rate coefficient does not play a significant role as long as the branching ratio is $\lesssim$~1:5. Furthermore, the H$_{3}^{+}$\,/\,H$_{2}$ ratios in the ``D'' model are not significantly affected by variations in the DR branching ratio, the H$_{3}^{+}$\,/\,H$_{2}$ values are similar to those predicted for the ``S'' model. 
%
{The key conclusion is that while the total DR rate coefficient does not play a significant role as long as the branching ratio is $\lesssim$~1:5, a DR branching ratio approaching unity for $(1,1)-$ and $(1,0)-$H$_3^+$ is necessary in order to reproduce the observed distributions. }
Furthermore, the H$_{3}^{+}$\,/\,H$_{2}$ ratios in the ``D'' model are not significantly affected by variations in the DR branching ratio, the H$_{3}^{+}$\,/\,H$_{2}$ values are similar to those predicted for the ``S'' model. 

\begin{figure}[!htb]
\centering
\includegraphics[width=0.5\textwidth, angle=90]{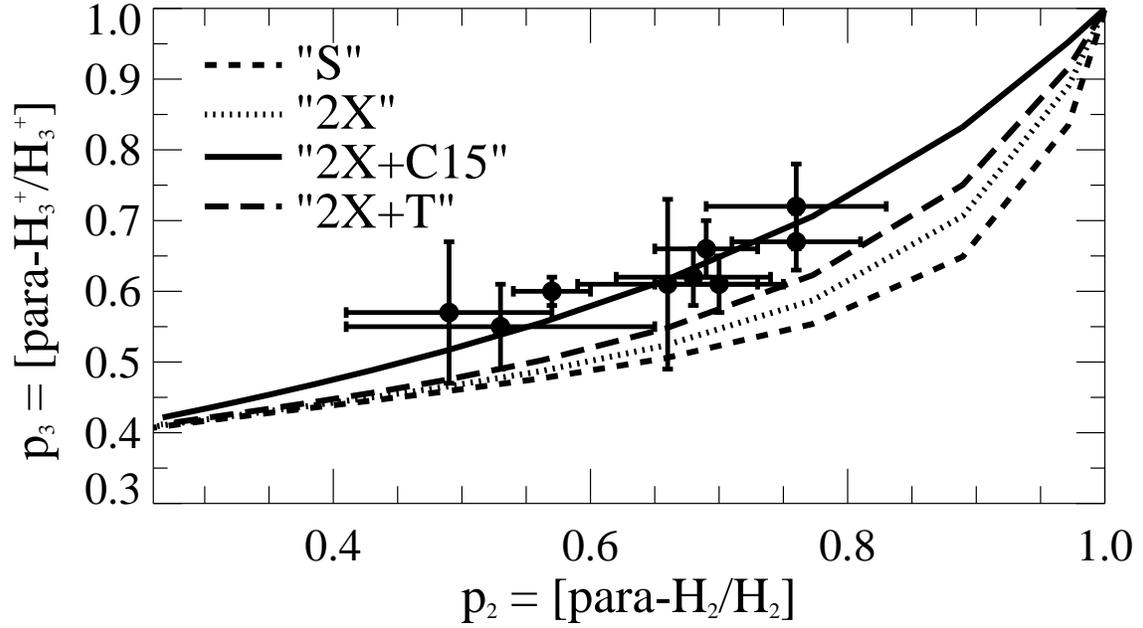} 
\caption{The \pHtre~and \pHtva~ values calculated with the standard chemical model for 1 Myr (dashed line) for temperatures of 10 - 100~K, with variations in the total DR reaction rates. DR rates are increased by a factor of 2 (dotted line, model ``2X'') {and also combined with an increase in chemical age to 10$^7$ years (long dashed line, model ``2X+T'') and an increase CR rate to $10^{-15}$ s$^{-1}$ (solid line, model ``2X+C15'')}. Observations are shown with error bars.}
\label{fig:DRtot}
\end{figure} 

It is clear that the increased DR rate of \textit{ortho}$-$H$_{3}^{+}$ improves the fit to observations, but the question remains if a higher total DR rate can also affect results, while the relative rate between $(1,1)-$ and $(1,0)-$H$_{3}^{+}$ remains the same. In Figure~\ref{fig:DRtot} we compare the results of increasing the total DR rate in the ``S'' model by a factor 2 (dotted line, model ``2X'') which significantly improves the fit. Because the $p_3$ value is largely determined by the ratio between thermalization and DR processes, an increase in the DR rate can also be considered the same effect as if the thermalization process is slowed down, as was discussed by \citet{2011ApJ...729...15C} (see their Figure~6). %It is likely that an increase by a factor of 4 is possible, if we consider an equivalent difference in the ratio between the thermalization and DR processes.

\subsection{Best-fit model}
A high total DR rate will cause the time effect to be more significant, hence we have also calculated a model at $10^{7}$ years. The additional longer time scale improves our fit {and we find that the resulting distribution of \textit{para}-fractions is similar to what is predicted in the ``2X'' model}. Furthermore, the total H$_{3}^{+}$ abundance decreases further, and at $10^{7}$ years H$_{3}^{+}$\,/\,H$_{2}$ =  $(1.41 - 2.62) \times 10^{-7}$, which is closer to the observed values. 

Because the H$_{3}^{+}$ abundance is also strongly affected, the H$_{3}^{+}$\,/\,H$_{2}$ ratio decreases by a similar factor of $\sim 2$ and $\sim 4$ for the same increase in the total DR rate, respectively. The H$_{3}^{+}$\,/\,H$_{2}$ ratio drops to values similar to the observed values of $\sim 10^{-7}$. It means that an increased total DR rate also has the added benefit of improving the agreement to the calculations of the H$_{3}^{+}$ column densities. {By combining the effects of an increased DR rate and increased chemical age (model ``2X + T'') we achieve a $p_3$ distribution that is approaching the observed distribution and where H$_{3}^{+}$\,/\,H$_{2}$ = $1.73 - 3.94 \times 10^{-7}$ is largely similar to the observed column densities. 
}

{A higher CR ionization rate (model ``C15'' with $\zeta_{CR}$ = 10$^{-15}$ s$^{-1}$) was shown to have a positive effect on improving the fit to observations. Combining the effect of the increased DR rate with an increased CR ionization rate (``2X + C15'') further helps to improve the agreement with observations. We find that the best fit to observations is achieved with the ``2X + C15'' model where predicted $p_2$ and $p_3$ values clearly go through the observation data points. The calculated H$_{3}^{+}$\,/\,H$_{2}$ = $4.61 - 7.62 \times 10^{-7}$ is somewhat higher due to the increased CR ionization rate, but similar to that predicted for the ``C15'' model. }

\subsection{Other molecules}\label{sec:molecules}
\begin{figure*}[!htb]
\centering
\includegraphics[width=0.45\textwidth, angle=90]{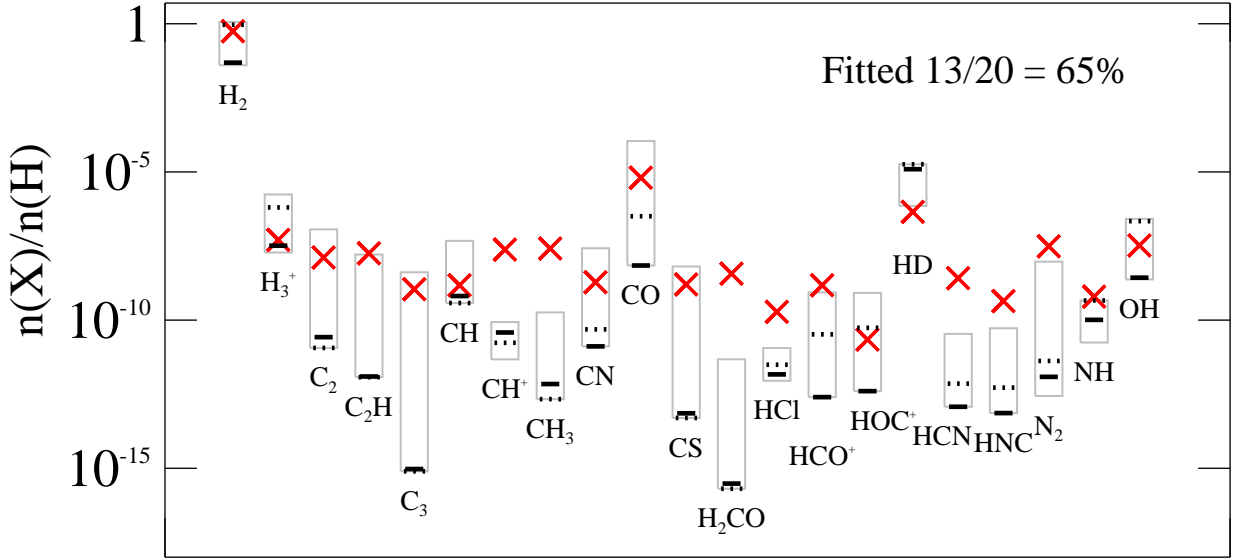} 
\caption{Comparison of observed abundances (red crosses) to modeled values of key species in diffuse clouds. Gray boxes show the range of abundances calculated from the considered models (Table~\ref{tab:models}) and black lines show abundances from the best-fit model ``2X+C15'' (30 K, solid line, and 90 K, dotted line). A colored version of the figure is available in the online version of the paper. }
\label{fig:abunds}
\end{figure*} 

{
While we have been concentrating on the abundances of H$_3^+$ and H$_2$, other molecules have also been observed in diffuse clouds and will be affected by the variations of model parameters. In Figure~\ref{fig:abunds}   the range of calculated abundances from the models in Table~\ref{tab:models} are compared to a compilation of observations in diffuse clouds from \citet{2006ARA&A..44..367S}. }

{
The calculated abundances cover a wide range that shows an agreement with several observed molecules (grey lines), especially  CH, CO, OH, H$_2$, H$_3^+$ and HD. Amongst the molecules difficult to fit we find molecules such as CH$^+$, HCN, HNC, HCO$^+$. If we consider only the best-fit model ``2X+C15'' we see that we underpredict abundances by many orders of magnitude (see gray lines of Figure~\ref{fig:abunds}), however the calculated H$_3^+$ abundances at $\sim 50$ K are in agreement with observed values. }

{
HCl shows a underprediction of abundances by approximately 2 orders of magnitude compared to observations, likely due to the very limited Cl-based chemistry in our network which makes our predicted HCl abundances very uncertain. 
}

Although CH$^+$ is a simple ion, it has for long been recognised as a problematic molecule to explain in the diffuse ISM \citep[see e.g.][]{1992MNRAS.255..463D, 1996MNRAS.279L..41F, 2009ApJ...692..594P}. Its expected slow formation through radiative association mechanism C$^+$ + H $\rightarrow$ CH$^+$ + $h\nu$ or C$^+$ + H$_2$ $\rightarrow$ CH$_2^+$ followed by photodissociation while being rapidly destroyed through reactions with H, H$_2$ and e$^-$ is  causing models to underpredict abundances by two orders of magnitude compared to observations. A possible solution to explain this discrepancy is finding a physical processes that would allow the endothermic reaction C$^+$ + H$_2$ $\rightarrow$ CH$^+$ + H to be effective and drive the production of CH$^+$, such as shocks or turbulence \citep[see e.g.][]{1986ApJ...310..392D, 2005A&A...433..997F}.

In the diffuse ISM, lacking efficient surface chemistry, H$_2$CO is formed through the reaction CH$_3$ + O $\rightarrow$ H$_2$CO + H. The underproduction of H$_2$CO should therefore be able to be traced through the CH$_3$ formation, which begins with CH$^+$ and the subsequent formation of CH$_2^+$ and CH$_3^+$ through hydrogenation by reacting with H$_2$. As we can see from Figure~\ref{fig:abunds}, CH$^+$ is similarly underproduced as CH$_3$ and H$_2$CO, and it is likely that the underproduction is related to that of CH$^+$, which is mainly formed through H$_3^+$ reacting with C atoms. Since we can show a good agreement to H$_3^+$ observations, we can conclude that C atoms must be underproduced in chemical models. However, because atomic C has not been observed, this can not be tested.

For species such as HCN, HNC and N$_2$, light hydrocarbons first need to be formed, making these species late-time and hence strongly dependent on the chemical age and calculated abundances are very uncertain. 

The abundances of these species may also be affected by other processes and parameters such as the clumpiness of the environments, initial abundances and the chemistry of molecules in excited states \citep[see e.g.][]{1996A&A...307..271S, 2003MNRAS.343.1257P, 2007ApJ...667..275B, 2010ApJ...713..662A, 2013ChRv..113.8738O, 2013ChRv..113.8710A}. 

\section{Conclusions}\label{sec:conclusions}
{We present the results of three new sight lines towards diffuse clouds, where both H$_{3}^{+}$ and H$_{2}$ have been observed in their \textit{ortho} and \textit{para} forms. The new observations follow the same trend as found by \citet{2011ApJ...729...15C}, lying between the nascent and thermalized distribution. We come to the conclusion that H$_{3}^{+}$ is not fully thermalized as the thermalization by collisions with H$_{2}$ is competing with destruction of H$_{3}^{+}$ by dissociative recombination.}

To study this, we conducted the first time-dependent modeling of the nuclear spin-states of H\dtwo~and H\dthree\jon~in the diffuse interstellar medium, and compared our results to the observed values, including the new measurements. 
%Our comprehensive time-dependent model {simple} explanations for the observed nuclear spin temperature discrepancy between H$_3^+$ and H$_2$.

We found that the DR of H$_3^+$ is a key process that governs the \pHtre~values. Our model indicates that a branching ratio of $\sim 1$ between the $(1,1)$- and $(1,0)$- H$_3^+$ dissociation is needed to achieve an agreement with the observations. {An increased CR ionization rate to 10$^{-15}$ s$^{-1}$ also has a significant effect on the $p_3$ values and brings the calculated values much closer to the observed values.} The remaining studied parameters, initial H$_2$ \textit{ortho}/\textit{para} ratio, $n_{H}$, chemical age and total DR rates will increase the pace at which the \pHtre~values approach the nascent distribution by a smaller, but significant amount. 

%Our model favors a gas hydrogen density of $\sim 10$ cm$^{-3}$ as a representative value for the diffuse ISM, in the lower range of diffuse ISM densities \citep{2006ARA&A..44..367S}. It allows the \pHtre~distribution to reach an agreement with the observed distribution within a reasonable time frame of $< 10^{8}$ years. Finally, we find that our predicted H$_{3}^{+}$\,/\,H$_{2}$ values match that derived from the observed column densities within an order of magnitude (where the largest impact is due to the adopted density and CR ionization rate).
%
{However, increasing the CR ionization rate to $10^{-15}$ s$^{-1}$ causes difficulties with reproducing observed abundances of other molecules, whereas many molecules are underproduced by our models. This is the same problem that has been raised before in the discussion of H$_3^+$ in diffuse clouds, but here we can also show that the high CR ionization rate is an essential ingredient in order to achieve an agreement with the \textit{para}-fractions of H$_3^+$. }

We conclude that the best fit to observations is achieved {for the ``2X+C15'' model with} a density of 10 cm$^{-3}$, CR ionization rate {10$^{-15}$ s$^{-1}$}, a 1:1 DR branching ratio, a time scale of {$10^6$} years and a total DR rate a factor of 2 larger than that derived by \citep{2004PhRvA..70e2716M}, or an equivalent reduction of thermalization rates \citep[see Figure 6 of][]{2011ApJ...729...15C}. With this model, we find that our predicted H$_{3}^{+}$\,/\,H$_{2}$ values match that derived from the observed column densities within an order of magnitude (where the largest impact is due to the adopted density and CR ionization rate). 

{These results warrant a more detailed study with a better treatment of the clumpy structure of diffuse clouds. Furthermore, better understanding of the CR ionization rate, and its possible variation within the diffuse cloud, and chemical ages need to be better constrained for future studies \citep[e.g.][]{2012A&A...537A...7R}.} It is {also} evident that the H$_3^+$ DR process is vital in order to better understand the \textit{ortho}$-$\textit{para} hydrogen chemistry in the diffuse ISM. For that, one has to bring in agreement the laboratory results on the DR of H$_3^+$ obtained with various experimental setups, such as the storage ring experiments \citep{kreckel05, tom09, kreckel10}, and the afterglow experiments \citep[e.g.][]{1995IJMSI.149..131G, 1998JPhB...31.2111L, 2002IJMSp.218..105P, glosik09}. Therefore, we highly recommend new accurate studies of the H$_{3}^{+}$ DR reactions, in order to both determine the absolute DR rate as well as the nuclear spin dependence for the lowest rotational states. 

\acknowledgments
We would like to thank the anonymous referee for their useful comments that helped improve this paper. This research made use of NASA's Astrophysics Data System. TA acknowledges funding from the European Community's Seventh Framework Programme [FP7/2007-2013] under grant agreement no. 238258. N.I. is funded by NASA Research Support Agreement No. 1465490 provided through JPL. H.K. was supported by the European Research Council under Grant Agreement No. StG 307163. DS acknowledges support by the {\it Deutsche Forschungsgemeinschaft} through SPP~1385: ``The first ten million years of the Solar System - a planetary materials approach'' (SE 1962/1-3). KNC has been supported by a CfA Postdoctoral Fellowship from the Smithsonian Astrophysical Observatory. 

\begin{deluxetable}{lclcccl}
\tablewidth{0pt}
\tabletypesize{\small}
\tablecaption{The dominant reactions for the hydrogen chemistry in the diffuse ISM. Values in parenthesis are exponential factors. Errors are defined as values for a log-normal distribution, with standard deviation $k \pm$ error. Errors marked with an asterisk (*) are resulting from our separation scheme to generate \textit{ortho}$-$\textit{para} reactions, and uncertainties are taken from the original reaction, but are likely higher. \label{tab:param}}
\tablehead{Reaction				&			&					&	{$\alpha$}		&	{$\beta$}		&	{$\gamma$}	& 	Error}
\startdata
H   +  CR   					&$\rightarrow$	& H$^+$   + e$^-$ 		& $0.46$		& $0$   & $0$   		& 2.00		\\
oH$_2$   +  CR   				&$\rightarrow$	& H$^+$   + H   + e$^-$   	& $0.02$   		& $0$   & $0$   		& 2.00$^*$		\\
pH$_2$   +  CR   				&$\rightarrow$	& H$^+$   + H   + e$^-$   	& $0.02$   		& $0$   & $0$   		& 2.00$^*$		\\
oH$_2$   +  CR	   				&$\rightarrow$	& oH$_2$$^+$   + e$^-$  	& $0.93$   		& $0$   & $0$   		& 2.00$^*$		\\
pH$_2$   +  CR   				&$\rightarrow$	& pH$_2$$^+$   + e$^-$   	& $0.93$   		& $0$   & $0$   		& 2.00$^*$		\\
oH$_2$   +  CR   				&$\rightarrow$	& 2H   				& $0.10$   		& $0$   & $0$   		& 2.00$^*$		\\
pH$_2$   +  CR	 				&$\rightarrow$	& 2H   	   			& $0.10$   		& $0$   & $0$   		& 2.00$^*$		\\
He   +  CR		   				&$\rightarrow$	& He$^+$   + e$^-$   		& $0.50$   		& $0$   & $0$   		& 2.00		\\
C   +  CR   					&$\rightarrow$	& C$^+$   + e$^-$   		& $1.02\,(3)$   	& $0$   & $0$   		& 2.00	\\
O   +  CR   					&$\rightarrow$	& O$^+$   + e$^-$   		& $2.80$   		& $0$   & $0$   		& 2.00		\\
S   + h$\nu_{\rm CR}$   			&$\rightarrow$	& S$^+$   + e$^-$   		& $9.60\,(2)$   	& $0$   & $0$   		& 2.00		\\
OH   + h$\nu_{\rm CR}$   			&$\rightarrow$	& O   + H   			& $5.10\,(2)$   	& $0$   & $0$   		& 2.00		\\
C   + h$\nu_{\rm CR}$   			&$\rightarrow$	& C$^+$   + e$^-$   		& $1.02\,(3)$   	& $0$   & $0$   		& 2.00		\\
H   + h$\nu_{\rm CR}$   			&$\rightarrow$	& H$^+$   + e$^-$   		& $0.46$   		& $0$   & $0$   		& 2.00		\\
He   + h$\nu_{\rm CR}$   			&$\rightarrow$	& He$^+$   + e$^-$   		& $0.50$   		& $0$   & $0$   		& 2.00		\\
O   + h$\nu_{\rm CR}$   			&$\rightarrow$	& O$^+$   + e$^-$   		& $2.80$   		& $0$   & $0$   		& 2.00		\\
oH$_2$   + h$\nu_{\rm CR}$   		&$\rightarrow$	& 2H  	  			& $0.10$   		& $0$   & $0$   		& 2.00$^*$		\\
pH$_2$   + h$\nu_{\rm CR}$   		&$\rightarrow$	& 2H   	   			& $0.10$   		& $0$   & $0$   		& 2.00$^*$		\\
oH$_2$   + h$\nu_{\rm CR}$   		&$\rightarrow$	& H$^+$   + H   + e$^-$   	& $0.02$   		& $0$   & $0$   		& 2.00$^*$		\\
pH$_2$   + h$\nu_{\rm CR}$   		&$\rightarrow$	& H$^+$   + H   + e$^-$   	& $0.02$   		& $0$   & $0$   		& 2.00$^*$		\\
oH$_2$   + h$\nu_{\rm CR}$   		&$\rightarrow$	& oH$_2$$^+$   + e$^-$   	& $0.93$   		& $0$   & $0$   		& 2.00$^*$		\\
pH$_2$   + h$\nu_{\rm CR}$   		&$\rightarrow$	& pH$_2$$^+$   + e$^-$   	& $0.93$   		& $0$   & $0$   		& 2.00$^*$		\\
C   + UV   						&$\rightarrow$	& C$^+$   + e$^-$   		& $0.22\,(-9)$   	& $0$   & $2.61$   	& 2.00		\\
OH   + UV   					&$\rightarrow$	& OH$^+$   + e$^-$   		& $0.16\,(-11)$   & $0$   & $3.10$   	& 2.00		\\
OH   + UV   					&$\rightarrow$	& O   + H   			& $0.17\,(-9)$   	& $0$   & $1.66$   	& 2.00		\\
OH$^+$   + UV   				&$\rightarrow$	& H$^+$   + O   			& $0.72\,(-11)$   & $0$   & $1.80$   	& 2.00		\\
S   + UV   						&$\rightarrow$	& S$^+$   + e$^-$  		& $0.72\,(-9)$   	& $0$   & $2.40$   	& 2.00		\\
oH$_2$   + UV   				&$\rightarrow$	& 2H   	   			& $0.34\,(-10)$  	& $0$   & $2.50$   	& 2.00$^*$		\\
pH$_2$   + UV   				&$\rightarrow$	& 2H   	   			& $0.34\,(-10)$  	& $0$   & $2.50$   	& 2.00$^*$		\\
oH$_2$$^+$   + UV   				&$\rightarrow$	& H$^+$   + H   			& $0.26\,(-9)$   	& $0$   & $1.80$   	& 2.00$^*$		\\
pH$_2$$^+$   + UV   				&$\rightarrow$	& H$^+$   + H   			& $0.26\,(-9)$   	& $0$   & $1.80$   	& 2.00$^*$		\\
C$^+$   + S   					&$\rightarrow$	& S$^+$   + C   			& $0.15\,(-8)$   	& $0$   & $0$   		& 2.00		\\
H$^+$   + O   					&$\rightarrow$	& O$^+$   + H   			& $0.70\,(-9)$   	& $0$   & $2.32\,(2)$  	& 1.50		\\
H$^+$   + OH   					&$\rightarrow$	& OH$^+$   + H   		& $0.16\,(-7)$   	& $-0.50$   & $0$   	& 2.00		\\
H$^+$   + S  	 				&$\rightarrow$	& S$^+$   + H   			& $0.13\,(-8)$   	& $0$   & $0$   		& 2.00		\\
oH$_2$$^+$   + H   				&$\rightarrow$	& H$^+$   + oH$_2$   	& $0.64\,(-9)$   	& $0$   & $0$   		& 1.25$^*$		\\
pH$_2$$^+$   + H   				&$\rightarrow$	& H$^+$   + pH$_2$   	& $0.64\,(-9)$   	& $0$   & $0$   		& 1.25$^*$		\\
pH$_2$$^+$   + oH$_2$   			&$\rightarrow$	& pH$_3$$^+$   + H   	& $0.14\,(-8)$   	& $0$   & $0$   		& 2.00		\\
oH$_2$$^+$   + O   				&$\rightarrow$	& OH$^+$   + H   		& $0.15\,(-8)$   	& $0$   & $0$   		& 2.00$^*$		\\
pH$_2$$^+$   + O   				&$\rightarrow$	& OH$^+$   + H   		& $0.15\,(-8)$   	& $0$   & $0$   		& 2.00$^*$		\\
oH$_2$$^+$   + OH   				&$\rightarrow$	& H$_2$O$^+$   + H   	& $0.76\,(-9)$   	& $0$   & $0$   		& 2.00$^*$		\\
pH$_2$$^+$   + OH   				&$\rightarrow$	& H$_2$O$^+$   + H   	& $0.76\,(-9)$   	& $0$   & $0$   		& 2.00$^*$		\\
oH$_2$$^+$   + OH   				&$\rightarrow$	& OH$^+$   + oH$_2$   	& $0.76\,(-9)$   	& $0$   & $0$   		& 2.00$^*$		\\
pH$_2$$^+$   + OH   				&$\rightarrow$	& OH$^+$   + pH$_2$   	& $0.76\,(-9)$   	& $0$   & $0$   		& 2.00$^*$		\\
oH$_3$$^+$   + O   				&$\rightarrow$	& OH$^+$   + oH$_2$   	& $0.80\,(-9)$   	& $-0.16$   & $1.41$  	& 1.41$^*$		\\
pH$_3$$^+$   + O   				&$\rightarrow$	& OH$^+$   + oH$_2$   	& $0.40\,(-9)$   	& $-0.16$   & $1.41$  	& 1.41$^*$		\\
pH$_3$$^+$   + O   				&$\rightarrow$	& OH$^+$   + pH$_2$   	& $0.40\,(-9)$   	& $-0.16$   & $1.41$  	& 1.41$^*$		\\
oH$_3$$^+$   + OH   				&$\rightarrow$	& H$_2$O$^+$   + oH$_2$ & $0.95\,(-8)$   	& $-0.50$   & $0$   	& 2.00$^*$		\\
pH$_3$$^+$   + OH   				&$\rightarrow$	& H$_2$O$^+$   + oH$_2$ & $0.47\,(-8)$   	& $-0.50$   & $0$   	& 2.00$^*$		\\
pH$_3$$^+$   + OH   				&$\rightarrow$	& H$_2$O$^+$   + pH$_2$ & $0.47\,(-8)$   	& $-0.50$   & $0$   	& 2.00$^*$		\\
He$^+$   + H   					&$\rightarrow$	& H$^+$   + He   		& $0.19\,(-14)$  	& $0$   & $0$   		& 1.25		\\
He$^+$   + oH$_2$   				&$\rightarrow$	& oH$_2$$^+$   + He   	& $0.96\,(-14)$  	& $0$   & $0$   		& 2.00$^*$		\\
He$^+$   + pH$_2$   				&$\rightarrow$	& oH$_2$$^+$   + He   	& $0.64\,(-14)$  	& $0$   & $0$   		& 2.00$^*$		\\
He$^+$   + pH$_2$   				&$\rightarrow$	& pH$_2$$^+$   + He   	& $0.32\,(-14)$  	& $0$   & $0$   		& 2.00$^*$		\\
He$^+$   + oH$_2$   				&$\rightarrow$	& H$^+$   + H   + He   	& $0.11\,(-12)$  	& $-0.24$   & $0$   	& 2.00$^*$		\\ 
He$^+$   + pH$_2$   				&$\rightarrow$	& H$^+$   + H   + He   	& $0.11\,(-12)$   & $-0.24$   & $0$   	& 2.00$^*$		\\ 
He$^+$   + OH   				&$\rightarrow$	& O$^+$   + H   + He   	& $0.85\,(-8)$   	& $-0.50$   & $0$   	& 2.00		\\
O$^+$   + H   					&$\rightarrow$	& H$^+$   + O   			& $0.70\,(-9)$   	& $0$   & $0$   		& 1.50		\\
O$^+$   + oH$_2$   				&$\rightarrow$	& OH$^+$   + H   		& $0.16\,(-8)$   	& $0$   & $0$   		& 1.25$^*$		\\
O$^+$   + pH$_2$   				&$\rightarrow$	& OH$^+$   + H   		& $0.16\,(-8)$   	& $0$   & $0$   		& 1.25$^*$		\\
O$^+$   + OH   					&$\rightarrow$	& OH$^+$   + O   		& $0.36\,(-9)$   	& $0$   & $0$   		& 2.00		\\
OH$^+$   + oH$_2$   				&$\rightarrow$	& H$_2$O$^+$   + H   	& $0.11\,(-8)$   	& $0$   & $0$   		& 1.25$^*$		\\
OH$^+$   + pH$_2$   				&$\rightarrow$	& H$_2$O$^+$   + H   	& $0.11\,(-8)$   	& $0$   & $0$   		& 1.25$^*$		\\
OH$^+$   + OH   				&$\rightarrow$	& H$_2$O$^+$   + O   	& $0.70\,(-9)$   	& $0$   & $0$   		& 2.00		\\
OH$^+$   + S   					&$\rightarrow$	& S$^+$   + OH   		& $0.43\,(-9)$   	& $0$   & $0$   		& 2.00		\\
H$^+$   + H   					&$\rightarrow$	& oH$_2$$^+$   		& $0.20\,(-19)$  	& $1.00$   & $0$  	& 2.00$^*$		\\
H$^+$   + H   					&$\rightarrow$	& pH$_2$$^+$   		& $0.20\,(-19)$  	& $1.00$   & $0$   	& 2.00$^*$		\\
H   + OH   						&$\rightarrow$	& O   + oH$_2$   		& $0.69\,(-13)$  	& $2.80$   & $1.70\,(2)$   & 2.00$^*$		\\
H   + OH   						&$\rightarrow$	& O   + pH$_2$   		& $0.69\,(-13)$  	& $2.80$   & $1.95\,(3)$   & 2.00$^*$		\\
oH$_2$   + O   					&$\rightarrow$	& OH   + H  	 		& $0.34\,(-12)$  	& $2.67$   & $3.16\,(3)$   & 3.16$^*$		\\
pH$_2$   + O   					&$\rightarrow$	& OH   + H   			& $0.34\,(-12)$  	& $2.67$   & $3.16\,(3)$   & 3.16$^*$		\\
H   + O   						&$\rightarrow$	& OH   				& $0.99\,(-18)$  	& $-0.38$   & $0$   	& 	10.0	\\
oH$_2$$^+$   + e$^-$   			&$\rightarrow$	& 2H  	   			& $0.16\,(-7)$   	& $-0.43$   & $0$   	& 	2.00$^*$	\\
pH$_2$$^+$   + e$^-$   			&$\rightarrow$	& 2H   	  			& $0.16\,(-7)$   	& $-0.43$   & $0$   	& 	2.00$^*$	\\
H$_2$O$^+$   + e$^-$   			&$\rightarrow$	& OH   + H   			& $0.86\,(-7)$   	& $-0.50$   & $0$   	& 	1.25	\\
H$_2$O$^+$   + e$^-$   			&$\rightarrow$	& O   + H   + H   		& $0.30\,(-6)$   	& $-0.50$   & $0$   	& 	1.25	\\
H$_2$O$^+$   + e$^-$   			&$\rightarrow$	& O   + oH$_2$   		& $0.39\,(-7)$   	& $-0.50$   & $1.70\,(2)$   & 1.25$^*$		\\
H$_2$O$^+$   + e$^-$   			&$\rightarrow$	& O   + pH$_2$   		& $0.39\,(-7)$   	& $-0.50$   & $0$   & 1.25$^*$		\\
oH$_3$$^+$   + e$^-$   			&$\rightarrow$	& oH$_2$   + H   		& \multicolumn{4}{c}{see Section~\ref{sec:H3dr}} 		\\
pH$_3$$^+$   + e$^-$   			&$\rightarrow$	& oH$_2$   + H   		& \multicolumn{4}{c}{see Section~\ref{sec:H3dr}}		\\
pH$_3$$^+$   + e$^-$   			&$\rightarrow$	& pH$_2$   + H   		& \multicolumn{4}{c}{see Section~\ref{sec:H3dr}}		\\
oH$_3$$^+$   + e$^-$   			&$\rightarrow$	& 3H 	  	 		& \multicolumn{4}{c}{see Section~\ref{sec:H3dr}}	\\
pH$_3$$^+$   + e$^-$   			&$\rightarrow$	& 3H   	  	 		& \multicolumn{4}{c}{see Section~\ref{sec:H3dr}}	\\
OH$^+$   + e$^-$  	 			&$\rightarrow$	& O   + H   			& $0.63\,(-8)$   	& $-0.48$   & $0$   		& 1.25		\\
C$^+$   + e$^-$   				&$\rightarrow$	& C   				& $0.44\,(-11)$   & $-0.61$   & $0$   		& 1.50		\\
H$^+$   + e$^-$   				&$\rightarrow$	& H   				& $0.35\,(-11)$   & $-0.70$   & $0$   		& 2.00	\\
He$^+$   + e$^-$   				&$\rightarrow$	& He   				& $0.45\,(-11)$   & $-0.67$   & $0$   		& 2.00	\\
O$^+$   + e$^-$   				&$\rightarrow$	& O   				& $0.34\,(-11)$   & $-0.63$   & $0$   		& 2.00		\\
S$^+$   + e$^-$   				&$\rightarrow$	& S   				& $0.39\,(-11)$   & $-0.63$   & $0$   		& 2.00		\\
oH$_3$$^+$   + O   				&$\rightarrow$	& H$_2$O$^+$   + H   	& $0.34\,(-9)$   	& $-0.16$   & $1.41$   	& 1.41$^*$		\\
pH$_3$$^+$   + O   				&$\rightarrow$	& H$_2$O$^+$   + H   	& $0.34\,(-9)$   	& $-0.16$   & $1.41$   	& 1.41$^*$		\\
H   + H   						&$\rightarrow$	& oH$_2$   			& $0.50\,(-16)$  	& $0.50$   & $1.70\,(2)$   	& 2.00$^*$		\\
H   + H   						&$\rightarrow$	& pH$_2$   			& $0.50\,(-16)$  	& $0.50$   & $0$   		& 2.00$^*$		\\
H$^+$   + oH$_2$   				&$\rightarrow$	& oH$_2$$^+$   + H   	& $0.64\,(-9)$   	& $0$   & $2.13\,(4)$   	& 2.00		\\
H$^+$   + pH$_2$   				&$\rightarrow$	& oH$_2$$^+$   + H   	& $0.43\,(-9)$   	& $0$   & $2.13\,(4)$   	& 2.00		\\
H$^+$   + pH$_2$   				&$\rightarrow$	& pH$_2$$^+$   + H   	& $0.21\,(-9)$   	& $0$   & $2.13\,(4)$   	& 2.00		\\
oH$_3$$^+$   + H   				&$\rightarrow$	& pH$_2$$^+$   + oH$_2$  & $0.21\,(-8)$   	& $0$   & $2.00\,(4)$   	& 2.00$^*$		\\
pH$_3$$^+$   + H  				&$\rightarrow$	& pH$_2$$^+$   + oH$_2$  & $0.10\,(-8)$   	& $0$   & $2.00\,(4)$   	& 2.00$^*$		\\
pH$_3$$^+$   + H   				&$\rightarrow$	& pH$_2$$^+$   + pH$_2$  & $0.10\,(-8)$   	& $0$   & $2.00\,(4)$   	& 2.00$^*$		\\
C$^+$   + H   					&$\rightarrow$	& H$^+$  + C  			& $0.93\,(-18)$  	& $1.30$   & $1.59\,(4)$   	& 2.00		\\
S$^+$   + H   					&$\rightarrow$	& H$^+$  + S  			& $0.57\,(-15)$  	& $1.20$   & $2.72\,(4)$   	& 2.00		\\
H$^+$   + C     					&$\rightarrow$	& C$^+$  + H   			& $1.00\,(-14)$  	& $0$   	& $0$   		& 2.00		\\
pH$_3$$^+$   + oH$_2$   			&$\rightarrow$	& pH$_3$$^+$   + 2H 	& $0.30\,(-10)$  	& $0.50$   & $5.20\,(4)$   	& 2.00		\\
He$^+$   + oH$_2$   				&$\rightarrow$	& He$^+$   + 2H 	   	& $0.30\,(-10)$  	& $0.50$   & $5.20\,(4)$   	& 2.00$^*$		
\enddata
\end{deluxetable}

\clearpage
\bibliography{OrthoParaDiffuseISM}{}

\begin{thebibliography}{85}
\expandafter\ifx\csname natexlab\endcsname\relax\def\natexlab#1{#1}\fi

\bibitem[{{Ag{\'u}ndez} {et~al.}(2010){Ag{\'u}ndez}, {Goicoechea},
  {Cernicharo}, {Faure}, \& {Roueff}}]{2010ApJ...713..662A}
{Ag{\'u}ndez}, M., {Goicoechea}, J.~R., {Cernicharo}, J., {Faure}, A., \&
  {Roueff}, E. 2010, \apj, 713, 662

\bibitem[{{Ag{\'u}ndez} \& {Wakelam}(2013)}]{2013ChRv..113.8710A}
{Ag{\'u}ndez}, M. \& {Wakelam}, V. 2013, Chemical Reviews, 113, 8710

\bibitem[{{Albertsson} {et~al.}(2013){Albertsson}, {Semenov}, {Vasyunin},
  {Henning}, \& {Herbst}}]{2013ApJS..207...27A}
{Albertsson}, T., {Semenov}, D.~A., {Vasyunin}, A.~I., {Henning}, T., \&
  {Herbst}, E. 2013, \apjs, 207, 27

\bibitem[{{Bethell} {et~al.}(2007){Bethell}, {Zweibel}, \&
  {Li}}]{2007ApJ...667..275B}
{Bethell}, T.~J., {Zweibel}, E.~G., \& {Li}, P.~S. 2007, \apj, 667, 275

\bibitem[{{Cardelli} {et~al.}(1996){Cardelli}, {Meyer}, {Jura}, \&
  {Savage}}]{1996ApJ...467..334C}
{Cardelli}, J.~A., {Meyer}, D.~M., {Jura}, M., \& {Savage}, B.~D. 1996, \apj,
  467, 334

\bibitem[{{Caselli}(2003)}]{2003Ap&SS.285..619C}
{Caselli}, P. 2003, \apss, 285, 619

\bibitem[{{Cecchi-Pestellini} {et~al.}(2009){Cecchi-Pestellini}, {Williams},
  {Viti}, \& {Casu}}]{2009ApJ...706.1429C}
{Cecchi-Pestellini}, C., {Williams}, D.~A., {Viti}, S., \& {Casu}, S. 2009,
  \apj, 706, 1429

\bibitem[{{Chehrouri} {et~al.}(2011){Chehrouri}, {Fillion}, {Chaabouni},
  {Mokrane}, {Congiu}, {Dulieu}, {Matar}, {Michaut}, \&
  {Lemaire}}]{2011PCCP...13.2172C}
{Chehrouri}, M., {Fillion}, J.-H., {Chaabouni}, H., {et~al.} 2011, Physical
  Chemistry Chemical Physics (Incorporating Faraday Transactions), 13, 2172

\bibitem[{{Crabtree} {et~al.}(2011){Crabtree}, {Indriolo}, {Kreckel}, {Tom}, \&
  {McCall}}]{2011ApJ...729...15C}
{Crabtree}, K.~N., {Indriolo}, N., {Kreckel}, H., {Tom}, B.~A., \& {McCall},
  B.~J. 2011, \apj, 729, 15

\bibitem[{Crabtree \& McCall(2012)}]{Crabtree13112012}
Crabtree, K.~N. \& McCall, B.~J. 2012, Philosophical Transactions of the Royal
  Society A: Mathematical, Physical and Engineering Sciences, 370, 5055

\bibitem[{Crabtree \& McCall(2013)}]{Crabtree2013}
Crabtree, K.~N. \& McCall, B.~J. 2013, The Journal of Physical Chemistry A,
  117, 9950

\bibitem[{dos Santos {et~al.}(2007)dos Santos, Kokoouline, \&
  Greene}]{santos:124309}
dos Santos, S.~F., Kokoouline, V., \& Greene, C.~H. 2007, The Journal of
  Chemical Physics, 127, 124309

\bibitem[{{Draine} \& {Bertoldi}(1996)}]{1996ApJ...468..269D}
{Draine}, B.~T. \& {Bertoldi}, F. 1996, \apj, 468, 269

\bibitem[{{Draine} \& {Katz}(1986)}]{1986ApJ...310..392D}
{Draine}, B.~T. \& {Katz}, N. 1986, \apj, 310, 392

\bibitem[{{Duley} {et~al.}(1992){Duley}, {Hartquist}, {Sternberg},
  {Wagenblast}, \& {Williams}}]{1992MNRAS.255..463D}
{Duley}, W.~W., {Hartquist}, T.~W., {Sternberg}, A., {Wagenblast}, R., \&
  {Williams}, D.~A. 1992, \mnras, 255, 463

\bibitem[{{Falgarone} {et~al.}(2013){Falgarone}, {Godard}, {Pineaue des
  For{\^e}ts}, \& {Gerin}}]{2013IAUS..292..223F}
{Falgarone}, E., {Godard}, B., {Pineaue des For{\^e}ts}, G., \& {Gerin}, M.
  2013, in IAU Symposium, Vol. 292, IAU Symposium, ed. T.~{Wong} \& J.~{Ott},
  223--226

\bibitem[{{Falgarone} {et~al.}(2005){Falgarone}, {Verstraete}, {Pineau Des
  For{\^e}ts}, \& {Hily-Blant}}]{2005A&A...433..997F}
{Falgarone}, E., {Verstraete}, L., {Pineau Des For{\^e}ts}, G., \&
  {Hily-Blant}, P. 2005, \aap, 433, 997

\bibitem[{{Federman} {et~al.}(1996){Federman}, {Rawlings}, {Taylor}, \&
  {Williams}}]{1996MNRAS.279L..41F}
{Federman}, S.~R., {Rawlings}, J.~M.~C., {Taylor}, S.~D., \& {Williams}, D.~A.
  1996, \mnras, 279, L41

\bibitem[{{Flower} {et~al.}(2004){Flower}, {Pineau des For{\^e}ts}, \&
  {Walmsley}}]{2004A&A...427..887F}
{Flower}, D.~R., {Pineau des For{\^e}ts}, G., \& {Walmsley}, C.~M. 2004, \aap,
  427, 887

\bibitem[{{Geballe} {et~al.}(1999){Geballe}, {McCall}, {Hinkle}, \&
  {Oka}}]{1999ApJ...510..251G}
{Geballe}, T.~R., {McCall}, B.~J., {Hinkle}, K.~H., \& {Oka}, T. 1999, \apj,
  510, 251

\bibitem[{{Geballe} \& {Oka}(1996)}]{1996Natur.384..334G}
{Geballe}, T.~R. \& {Oka}, T. 1996, \nat, 384, 334

\bibitem[{{Gerlich}(1990)}]{1990JChPh..92.2377G}
{Gerlich}, D. 1990, \jcp, 92, 2377

\bibitem[{{Gibb} {et~al.}(2010){Gibb}, {Brittain}, {Rettig}, {Troutman},
  {Simon}, \& {Kulesa}}]{2010ApJ...715..757G}
{Gibb}, E.~L., {Brittain}, S.~D., {Rettig}, T.~W., {et~al.} 2010, \apj, 715,
  757

\bibitem[{{Glos{\'{\i}}k} {et~al.}(2009){Glos{\'{\i}}k}, {Pla{\v s}il},
  {Korolov}, {Kotr{\'{\i}}k}, {Novotn{\'y}}, {Hlavenka}, {Dohnal}, {Varju},
  {Kokoouline}, \& {Greene}}]{glosik09}
{Glos{\'{\i}}k}, J., {Pla{\v s}il}, R., {Korolov}, I., {et~al.} 2009, \pra, 79,
  052707

\bibitem[{{Goto} {et~al.}(2008){Goto}, {Usuda}, {Nagata}, {Geballe}, {McCall},
  {Indriolo}, {Suto}, {Henning}, {Morong}, \& {Oka}}]{2008ApJ...688..306G}
{Goto}, M., {Usuda}, T., {Nagata}, T., {et~al.} 2008, \apj, 688, 306

\bibitem[{{Gougousi} {et~al.}(1995){Gougousi}, {Johnsen}, \&
  {Golde}}]{1995IJMSI.149..131G}
{Gougousi}, T., {Johnsen}, R., \& {Golde}, M.~F. 1995, International Journal of
  Mass Spectrometry and Ion Processes, 149, 131

\bibitem[{{Graedel} {et~al.}(1982){Graedel}, {Langer}, \&
  {Frerking}}]{1982ApJS...48..321G}
{Graedel}, T.~E., {Langer}, W.~D., \& {Frerking}, M.~A. 1982, \apjs, 48, 321

\bibitem[{{Grussie} {et~al.}(2012){Grussie}, {Berg}, {Crabtree}, {G{\"a}rtner},
  {McCall}, {Schlemmer}, {Wolf}, \& {Kreckel}}]{2012ApJ...759...21G}
{Grussie}, F., {Berg}, M.~H., {Crabtree}, K.~N., {et~al.} 2012, \apj, 759, 21

\bibitem[{{Honvault} {et~al.}(2011){Honvault}, {Jorfi}, {Gonz{\'a}lez-Lezana},
  {Faure}, \& {Pagani}}]{2011PhRvL.107b3201H}
{Honvault}, P., {Jorfi}, M., {Gonz{\'a}lez-Lezana}, T., {Faure}, A., \&
  {Pagani}, L. 2011, Physical Review Letters, 107, 023201

\bibitem[{{Hugo} {et~al.}(2009){Hugo}, {Asvany}, \&
  {Schlemmer}}]{2009JChPh.130p4302H}
{Hugo}, E., {Asvany}, O., \& {Schlemmer}, S. 2009, \jcp, 130, 164302

\bibitem[{{Indriolo} {et~al.}(2009){Indriolo}, {Fields}, \&
  {McCall}}]{2009ApJ...694..257I}
{Indriolo}, N., {Fields}, B.~D., \& {McCall}, B.~J. 2009, \apj, 694, 257

\bibitem[{{Indriolo} {et~al.}(2007){Indriolo}, {Geballe}, {Oka}, \&
  {McCall}}]{2007ApJ...671.1736I}
{Indriolo}, N., {Geballe}, T.~R., {Oka}, T., \& {McCall}, B.~J. 2007, \apj,
  671, 1736

\bibitem[{{Indriolo} \& {McCall}(2012)}]{2012ApJ...745...91I}
{Indriolo}, N. \& {McCall}, B.~J. 2012, \apj, 745, 91

\bibitem[{{Jenkins}(2009)}]{2009ApJ...700.1299J}
{Jenkins}, E.~B. 2009, \apj, 700, 1299

\bibitem[{{Kaeufl} {et~al.}(2004){Kaeufl}, {Ballester}, {Biereichel},
  {Delabre}, {Donaldson}, {Dorn}, {Fedrigo}, {Finger}, {Fischer}, {Franza},
  {Gojak}, {Huster}, {Jung}, {Lizon}, {Mehrgan}, {Meyer}, {Moorwood}, {Pirard},
  {Paufique}, {Pozna}, {Siebenmorgen}, {Silber}, {Stegmeier}, \&
  {Wegerer}}]{2004SPIE.5492.1218K}
{Kaeufl}, H.-U., {Ballester}, P., {Biereichel}, P., {et~al.} 2004, in Society
  of Photo-Optical Instrumentation Engineers (SPIE) Conference Series, Vol.
  5492, Society of Photo-Optical Instrumentation Engineers (SPIE) Conference
  Series, ed. A.~F.~M. {Moorwood} \& M.~{Iye}, 1218--1227

\bibitem[{{Kokoouline} \& {Greene}(2003{\natexlab{a}})}]{kokoouline03}
{Kokoouline}, V. \& {Greene}, C.~H. 2003{\natexlab{a}}, Physical Review
  Letters, 90, 133201

\bibitem[{{Kokoouline} \& {Greene}(2003{\natexlab{b}})}]{kokoouline03b}
{Kokoouline}, V. \& {Greene}, C.~H. 2003{\natexlab{b}}, \pra, 68, 012703

\bibitem[{{Kokoouline} {et~al.}(2001){Kokoouline}, {Greene}, \&
  {Esry}}]{kokoouline01}
{Kokoouline}, V., {Greene}, C.~H., \& {Esry}, B.~D. 2001, \nat, 412, 891

\bibitem[{{Kreckel} {et~al.}(2005){Kreckel}, {Motsch}, {Mikosch},
  {Glos{\'{\i}}k}, {Pla{\v s}il}, {Altevogt}, {Andrianarijaona}, {Buhr},
  {Hoffmann}, {Lammich}, {Lestinsky}, {Nevo}, {Novotny}, {Orlov}, {Pedersen},
  {Sprenger}, {Terekhov}, {Toker}, {Wester}, {Gerlich}, {Schwalm}, {Wolf}, \&
  {Zajfman}}]{kreckel05}
{Kreckel}, H., {Motsch}, M., {Mikosch}, J., {et~al.} 2005, Physical Review
  Letters, 95, 263201

\bibitem[{{Kreckel} {et~al.}(2010){Kreckel}, {Novotn{\'y}}, {Crabtree}, {Buhr},
  {Petrignani}, {Tom}, {Thomas}, {Berg}, {Bing}, {Grieser}, {Krantz},
  {Lestinsky}, {Mendes}, {Nordhorn}, {Repnow}, {St{\"u}tzel}, {Wolf}, \&
  {McCall}}]{kreckel10}
{Kreckel}, H., {Novotn{\'y}}, O., {Crabtree}, K.~N., {et~al.} 2010, \pra, 82,
  042715

\bibitem[{{Larsson}(2000)}]{larsson00}
{Larsson}, M. 2000, in Phil. Trans. R. Soc. Lond. A, Vol. 358, Astronomy,
  physics and chemistry of H$^{+}$$_{3}$, 2433--2444

\bibitem[{{Laub{\'e}} {et~al.}(1998){Laub{\'e}}, {LePadellec}, {Sidko},
  {Rebrion-Rowe}, {Mitchell}, \& {Rowe}}]{1998JPhB...31.2111L}
{Laub{\'e}}, S., {LePadellec}, A., {Sidko}, O., {et~al.} 1998, Journal of
  Physics B Atomic Molecular Physics, 31, 2111

\bibitem[{{Le Bourlot} {et~al.}(1995){Le Bourlot}, {Pineau des Forets}, \&
  {Roueff}}]{1995A&A...297..251L}
{Le Bourlot}, J., {Pineau des Forets}, G., \& {Roueff}, E. 1995, \aap, 297, 251

\bibitem[{{Lee} {et~al.}(1996){Lee}, {Herbst}, {Pineau des Forets}, {Roueff},
  \& {Le Bourlot}}]{1996A&A...311..690L}
{Lee}, H.-H., {Herbst}, E., {Pineau des Forets}, G., {Roueff}, E., \& {Le
  Bourlot}, J. 1996, \aap, 311, 690

\bibitem[{{Lee} {et~al.}(1998){Lee}, {Roueff}, {Pineau des Forets},
  {Shalabiea}, {Terzieva}, \& {Herbst}}]{1998A&A...334.1047L}
{Lee}, H.-H., {Roueff}, E., {Pineau des Forets}, G., {et~al.} 1998, \aap, 334,
  1047

\bibitem[{{Lindsay} \& {McCall}(2001)}]{2001JMoSp.210...60L}
{Lindsay}, C.~M. \& {McCall}, B.~J. 2001, Journal of Molecular Spectroscopy,
  210, 60

\bibitem[{{Liszt}(2006)}]{2006RSPTA.364.3049L}
{Liszt}, H.~S. 2006, Royal Society of London Philosophical Transactions Series
  A, 364, 3049

\bibitem[{{McCall} {et~al.}(1998){McCall}, {Geballe}, {Hinkle}, \&
  {Oka}}]{1998Sci...279.1910M}
{McCall}, B.~J., {Geballe}, T.~R., {Hinkle}, K.~H., \& {Oka}, T. 1998, Science,
  279, 1910

\bibitem[{{McCall} {et~al.}(2002){McCall}, {Hinkle}, {Geballe},
  {Moriarty-Schieven}, {Evans}, {Kawaguchi}, {Takano}, {Smith}, \&
  {Oka}}]{2002ApJ...567..391M}
{McCall}, B.~J., {Hinkle}, K.~H., {Geballe}, T.~R., {et~al.} 2002, \apj, 567,
  391

\bibitem[{{McCall} {et~al.}(2004){McCall}, {Huneycutt}, {Saykally}, {Djuric},
  {Dunn}, {Semaniak}, {Novotny}, {Al-Khalili}, {Ehlerding}, {Hellberg},
  {Kalhori}, {Neau}, {Thomas}, {Paal}, {{\"O}sterdahl}, \&
  {Larsson}}]{2004PhRvA..70e2716M}
{McCall}, B.~J., {Huneycutt}, A.~J., {Saykally}, R.~J., {et~al.} 2004, \pra,
  70, 052716

\bibitem[{{McCall} {et~al.}(2003){McCall}, {Huneycutt}, {Saykally}, {Geballe},
  {Djuric}, {Dunn}, {Semaniak}, {Novotny}, {Al-Khalili}, {Ehlerding},
  {Hellberg}, {Kalhori}, {Neau}, {Thomas}, {{\"O}sterdahl}, \&
  {Larsson}}]{2003Natur.422..500M}
{McCall}, B.~J., {Huneycutt}, A.~J., {Saykally}, R.~J., {et~al.} 2003, \nat,
  422, 500

\bibitem[{{McElroy} {et~al.}(2013){McElroy}, {Walsh}, {Markwick}, {Cordiner},
  {Smith}, \& {Millar}}]{2013A&A...550A..36M}
{McElroy}, D., {Walsh}, C., {Markwick}, A.~J., {et~al.} 2013, \aap, 550, A36

\bibitem[{{McKee} \& {Ostriker}(2007)}]{2007ARA&A..45..565M}
{McKee}, C.~F. \& {Ostriker}, E.~C. 2007, \araa, 45, 565

\bibitem[{{{\"O}berg} {et~al.}(2009{\natexlab{a}}){{\"O}berg}, {Linnartz},
  {Visser}, \& {van Dishoeck}}]{2009ApJ...693.1209O}
{{\"O}berg}, K.~I., {Linnartz}, H., {Visser}, R., \& {van Dishoeck}, E.~F.
  2009{\natexlab{a}}, \apj, 693, 1209

\bibitem[{{{\"O}berg} {et~al.}(2009{\natexlab{b}}){{\"O}berg}, {van Dishoeck},
  \& {Linnartz}}]{2009A&A...496..281O}
{{\"O}berg}, K.~I., {van Dishoeck}, E.~F., \& {Linnartz}, H.
  2009{\natexlab{b}}, \aap, 496, 281

\bibitem[{{Oka}(2004)}]{2004JMoSp.228..635O}
{Oka}, T. 2004, Journal of Molecular Spectroscopy, 228, 635

\bibitem[{{Oka}(2013)}]{2013ChRv..113.8738O}
{Oka}, T. 2013, Chemical Reviews, 113, 8738

\bibitem[{Okumura {et~al.}(2013)Okumura, McCall, \& Geballe}]{OkaFestschrift}
Okumura, M., McCall, B.~J., \& Geballe, T.~R. 2013, The Journal of Physical
  Chemistry A, 117, 9305

\bibitem[{{Pagani} {et~al.}(2011){Pagani}, {Roueff}, \&
  {Lesaffre}}]{2011ApJ...739L..35P}
{Pagani}, L., {Roueff}, E., \& {Lesaffre}, P. 2011, \apjl, 739, L35

\bibitem[{{Pagani} {et~al.}(2009){Pagani}, {Vastel}, {Hugo}, {Kokoouline},
  {Greene}, {Bacmann}, {Bayet}, {Ceccarelli}, {Peng}, \&
  {Schlemmer}}]{2009A&A...494..623P}
{Pagani}, L., {Vastel}, C., {Hugo}, E., {et~al.} 2009, \aap, 494, 623

\bibitem[{{Pan} \& {Padoan}(2009)}]{2009ApJ...692..594P}
{Pan}, L. \& {Padoan}, P. 2009, \apj, 692, 594

\bibitem[{{Petrignani} {et~al.}(2011){Petrignani}, {Altevogt}, {Berg}, {Bing},
  {Grieser}, {Hoffmann}, {Jordon-Thaden}, {Krantz}, {Mendes}, {Novotn{\'y}},
  {Novotny}, {Orlov}, {Repnow}, {Sorg}, {St{\"u}tzel}, {Wolf}, {Buhr},
  {Kreckel}, {Kokoouline}, \& {Greene}}]{petrignani11}
{Petrignani}, A., {Altevogt}, S., {Berg}, M.~H., {et~al.} 2011, \pra, 83,
  032711

\bibitem[{{Pla{\v s}il} {et~al.}(2002){Pla{\v s}il}, {Glos{\'{\i}}k},
  {Poterya}, {Kudrna}, {Rusz}, {Tich{\'y}}, \&
  {Pysanenko}}]{2002IJMSp.218..105P}
{Pla{\v s}il}, R., {Glos{\'{\i}}k}, J., {Poterya}, V., {et~al.} 2002,
  International Journal of Mass Spectrometry, 218, 105

\bibitem[{{Price} {et~al.}(2003){Price}, {Viti}, \&
  {Williams}}]{2003MNRAS.343.1257P}
{Price}, R.~J., {Viti}, S., \& {Williams}, D.~A. 2003, \mnras, 343, 1257

\bibitem[{{Quack}(1977)}]{1977MolPh..34..477Q}
{Quack}, M. 1977, Molecular Physics, 34, 477

\bibitem[{{Rachford} {et~al.}(2009){Rachford}, {Snow}, {Destree}, {Ross},
  {Ferlet}, {Friedman}, {Gry}, {Jenkins}, {Morton}, {Savage}, {Shull},
  {Sonnentrucker}, {Tumlinson}, {Vidal-Madjar}, {Welty}, \&
  {York}}]{2009ApJS..180..125R}
{Rachford}, B.~L., {Snow}, T.~P., {Destree}, J.~D., {et~al.} 2009, \apjs, 180,
  125

\bibitem[{{Rachford} {et~al.}(2002){Rachford}, {Snow}, {Tumlinson}, {Shull},
  {Blair}, {Ferlet}, {Friedman}, {Gry}, {Jenkins}, {Morton}, {Savage},
  {Sonnentrucker}, {Vidal-Madjar}, {Welty}, \& {York}}]{2002ApJ...577..221R}
{Rachford}, B.~L., {Snow}, T.~P., {Tumlinson}, J., {et~al.} 2002, \apj, 577,
  221

\bibitem[{{Rimmer} {et~al.}(2012){Rimmer}, {Herbst}, {Morata}, \&
  {Roueff}}]{2012A&A...537A...7R}
{Rimmer}, P.~B., {Herbst}, E., {Morata}, O., \& {Roueff}, E. 2012, \aap, 537,
  A7

\bibitem[{{Savage} {et~al.}(1977){Savage}, {Bohlin}, {Drake}, \&
  {Budich}}]{1977ApJ...216..291S}
{Savage}, B.~D., {Bohlin}, R.~C., {Drake}, J.~F., \& {Budich}, W. 1977, \apj,
  216, 291

\bibitem[{{Semenov} {et~al.}(2010){Semenov}, {Hersant}, {Wakelam}, {Dutrey},
  {Chapillon}, {Guilloteau}, {Henning}, {Launhardt}, {Pi{\'e}tu}, \&
  {Schreyer}}]{2010A&A...522A..42S}
{Semenov}, D., {Hersant}, F., {Wakelam}, V., {et~al.} 2010, \aap, 522, A42

\bibitem[{{Semenov} {et~al.}(2006){Semenov}, {Wiebe}, \&
  {Henning}}]{2006ApJ...647L..57S}
{Semenov}, D., {Wiebe}, D., \& {Henning}, T. 2006, \apjl, 647, L57

\bibitem[{{Sipil{\"a}} {et~al.}(2013){Sipil{\"a}}, {Caselli}, \&
  {Harju}}]{2013A&A...554A..92S}
{Sipil{\"a}}, O., {Caselli}, P., \& {Harju}, J. 2013, \aap, 554, A92

\bibitem[{{Snow} \& {McCall}(2006)}]{2006ARA&A..44..367S}
{Snow}, T.~P. \& {McCall}, B.~J. 2006, \araa, 44, 367

\bibitem[{{Spaans}(1996)}]{1996A&A...307..271S}
{Spaans}, M. 1996, \aap, 307, 271

\bibitem[{{Sternberg} \& {Neufeld}(1999)}]{1999ApJ...516..371S}
{Sternberg}, A. \& {Neufeld}, D.~A. 1999, \apj, 516, 371

\bibitem[{{Talbi} \& {Saxon}(1988)}]{1988JChPh..89.2235T}
{Talbi}, D. \& {Saxon}, R.~P. 1988, \jcp, 89, 2235

\bibitem[{{Tom} {et~al.}(2009){Tom}, {Zhaunerchyk}, {Wiczer}, {Mills},
  {Crabtree}, {Kaminska}, {Geppert}, {Hamberg}, {Af Ugglas}, {Vigren}, {van der
  Zande}, {Larsson}, {Thomas}, \& {McCall}}]{tom09}
{Tom}, B.~A., {Zhaunerchyk}, V., {Wiczer}, M.~B., {et~al.} 2009, \jcp, 130,
  031101

\bibitem[{{van der Tak}(2006)}]{2006RSPTA.364.3101V}
{van der Tak}, F.~F.~S. 2006, Royal Society of London Philosophical
  Transactions Series A, 364, 3101

\bibitem[{{van der Tak} \& {van Dishoeck}(2000)}]{2000A&A...358L..79V}
{van der Tak}, F.~F.~S. \& {van Dishoeck}, E.~F. 2000, \aap, 358, L79

\bibitem[{{van Dishoeck}(1987)}]{1987IAUS..120...51V}
{van Dishoeck}, E.~F. 1987, in IAU Symposium, Vol. 120, Astrochemistry, ed.
  M.~S. {Vardya} \& S.~P. {Tarafdar}, 51--63

\bibitem[{{van Dishoeck} {et~al.}(2006){van Dishoeck}, {Jonkheid}, \& {van
  Hemert}}]{2006FaDi..133..231V}
{van Dishoeck}, E.~F., {Jonkheid}, B., \& {van Hemert}, M.~C. 2006, Faraday
  Discussions, 133, 231

\bibitem[{{Varju} {et~al.}(2011){Varju}, {Hejduk}, {Dohnal}, {J{\'{\i}}lek},
  {Kotr{\'{\i}}k}, {Pla{\v s}il}, {Gerlich}, \&
  {Glos{\'{\i}}k}}]{2011PhRvL.106t3201V}
{Varju}, J., {Hejduk}, M., {Dohnal}, P., {et~al.} 2011, Physical Review
  Letters, 106, 203201

\bibitem[{{Walmsley} {et~al.}(2004){Walmsley}, {Flower}, \& {Pineau des
  For{\^e}ts}}]{2004A&A...418.1035W}
{Walmsley}, C.~M., {Flower}, D.~R., \& {Pineau des For{\^e}ts}, G. 2004, \aap,
  418, 1035

\bibitem[{{Watanabe} {et~al.}(2010){Watanabe}, {Kimura}, {Kouchi}, {Chigai},
  {Hama}, \& {Pirronello}}]{2010ApJ...714L.233W}
{Watanabe}, N., {Kimura}, Y., {Kouchi}, A., {et~al.} 2010, \apjl, 714, L233

\bibitem[{{Webber}(1998)}]{1998ApJ...506..329W}
{Webber}, W.~R. 1998, \apj, 506, 329

\end{thebibliography}
\bibliographystyle{aa}

\end{document}